\documentclass[a4paper,11pt]{article}
\pdfoutput=1 

\usepackage{jcappub} 

\usepackage[T1]{fontenc} 

\usepackage{graphics}
\usepackage{txfonts}

\usepackage{graphicx}
\usepackage{fixltx2e}
\usepackage{subfigure}
\usepackage{times}
\usepackage{amsmath}
\usepackage{amsfonts}
\usepackage{url}
\usepackage{placeins}
\usepackage{natbib}
\usepackage{amssymb,amsmath}
\usepackage[normalem]{ulem}
\usepackage{soul}
\usepackage{longtable}
\usepackage{pdflscape}
\usepackage{amsmath}
\usepackage{multirow}

\usepackage{lipsum}
\newcommand{\mhbar}{\bar{M}_h}
\newcommand{\mhmin}{M_{h,\rm min}}

\usepackage[normalem]{ulem}
\usepackage{soul}
\newcommand{\fduty}{f_{\rm duty}}

 \title{\boldmath Clustering of quasars in SDSS-IV eBOSS : study of potential systematics and bias determination.}

\author[a]{Pierre Laurent,}
\author[b]{Sarah Eftekharzadeh,}
\author[a,1]{Jean-Marc Le Goff,\note{Corresponding author.}}
\author[b]{Adam Myers,}
\author[a]{Etienne Burtin,}
\author[c,d,e]{Martin White,}

\author[f]{Ashley Ross,}
\author[g]{Jeremy Tinker,}
\author[h]{Rita Tojeiro,}

\author[i]{Julian Bautista,}
\author[j]{Johan Comparat,}
\author[a]{H\'elion du Mas des Bourboux,}
\author[k]{Jean-Paul Kneib,}
\author[l]{Ian D. McGreer,}
\author[a]{Nathalie Palanque-Delabrouille,}
\author[m]{Will J. Percival,}
\author[n]{Francisco Prada,}
\author[o]{Graziano Rossi,}
\author[p]{Donald P. Schneider,}
\author[q]{Micheal Strauss,}
\author[r]{David Weinberg,}
\author[a]{Christophe Y\`{e}che ,}
\author[a]{Pauline Zarrouk,}
\author[s]{and Gong-Bo Zhao}

\affiliation[a]{CEA, Centre de Saclay, IRFU/SPP,  F-91191 Gif-sur-Yvette, France}
\affiliation[b]{Department of Physics and Astronomy, University of Wyoming, Laramie, WY 82071, USA}
\affiliation[c]{Lawrence Berkeley National Lab, 1 Cyclotron Rd, Berkeley CA 94720, USA}
\affiliation[d]{Department of Physics, University of California, Berkeley, CA 94720, USA}
\affiliation[e]{Department of Astronomy, University of California, Berkeley, CA 94720, USA}

\affiliation[f]{Department of Astronomy, The Ohio State University, Columbus, OH 43210, USA}
\affiliation[g]{Center for Cosmology and Particle Physics, Department of Physics, New York University, New York, 10003, USA}
\affiliation[h]{School of Physics and Astronomy, North Haugh, St. Andrews KY16 9SS, UK}

\affiliation[i]{Department of Physics and Astronomy, University of Utah, Salt Lake City, UT 84112, USA}
\affiliation[j]{Max-Planck-Institut f\"{u}r Extraterrestrische Physik, Giessenbachstra{\ss}e, 85748 Garching, Germany}
\affiliation[k]{Laboratoire d'Astrophysique, \'{E}cole Polytechnique F\'{e}d\'{e}rale de Lausanne, 1015 Lausanne, Switzerland}
\affiliation[l]{Steward Observatory, University of Arizona, Tucson, AZ 85721–0065, USA}
\affiliation[m]{Institute of Cosmology and Gravitation, University of Portsmouth, Dennis Sciama building, PO1 3FX, Portsmouth, UK}
\affiliation[n]{Instituto de F\`{i}sica Te\'{o}rica (IFT) UAM/CSIC, Universidad Aut\'{o}noma de Madrid, Cantoblanco, E-28049 Madrid, Spain}
\affiliation[o]{Department of Astronomy and Space Science, Sejong University, Seoul, 143-747, Korea}
\affiliation[p]{Institute for Gravitation and the Cosmos, Pennsylvania State University, University Park, PA 16802, USA}
\affiliation[q]{Department of Astrophysical Sciences, Princeton, NJ 08544, USA}
\affiliation[r]{Department of Physics and Center for Cosmology and AstroParticle Physics, The Ohio State University, Columbus, OH 43210, USA}
\affiliation[s]{National Astronomical Observatories of China, Beijing 100012, China}

\emailAdd{jmlegoff@cea.fr}

\date{Received xx xx 2016 / accepted  xx xx 2016}

\abstract{We study the first year of the eBOSS quasar sample in the redshift range $0.9<z<2.2$ which includes 68,772 homogeneously selected quasars.
We show that the main source of systematics in the evaluation of the correlation function arises from inhomogeneities in the quasar target selection, particularly related to the extinction and depth of the imaging data used for targeting. We propose a weighting scheme that mitigates these systematics. We measure the quasar correlation function and provide the most accurate measurement to date of the quasar bias in this redshift range, $b_Q = 2.45 \pm 0.05$ at $\bar z=1.55$, together with its evolution with redshift. We use this information to determine the minimum mass of the halo hosting the quasars and the characteristic halo mass, which we find to be both independent of redshift within statistical error. Using a recently-measured quasar-luminosity-function we also determine the quasar duty cycle. The size of this first year sample is insufficient to detect any luminosity dependence to quasar clustering and this issue should be further studied with the final $\sim$500,000 eBOSS quasar sample.
}

\keywords{Large scale structure of the universe, redshift surveys, galaxy clustering}

\arxivnumber{}

\begin{document}
\toccontinuoustrue
\maketitle
\flushbottom

\section{Introduction}


The best current constraints on the cosmological parameters are from the
power spectrum of temperature fluctuations \citep[e.g.][]{Smo92,Ben96,Ben03,Kom09,Pla14} in the
Cosmic Microwave Background (henceforth CMB). In this regard, the latest Planck satellite results
provide overwhelming evidence for non-zero cosmic acceleration or ``dark energy''
with $\Omega_{\rm DE} = 0.692 \pm 0.012$ \citep{Pla15}. The CMB, however, can only provide a measurement at
one redshift (the epoch of the surface of last scattering at $z\sim1100$), and, so, measurements across many redshifts are required to
constrain the equation of state of dark energy \citep[e.g.][]{Che01,Lin05,Wei13}. As galaxies and quasars occupy a three-dimensional web that traces a range of
redshifts, they offer a unique probe of the evolution of dark energy over more than 10\,billion years of cosmic history.
Cosmological experiments are thus, increasingly, turning in part to vast redshift surveys in an attempt to map
the Universe in order to specifically study dark energy through the growth of structure \citep[e.g.][]{Dri10,Ada11,Daw13,Lev13,Dawson+2016}.

The first significant galaxy redshift surveys \citep[e.g.][]{Huc83,She96,Sau00}, were improved upon 
by surveys such as the 2dF Galaxy Redshift survey \citep{Col01}, the DEEP2 survey \citep{New13} and the Sloan Digital Sky Survey
\citep[henceforth SDSS;][]{Yor00} ``main'' galaxy sample \citep{Str02} and Luminous Red Galaxy (henceforth LRG)
sample \citep[][]{Eis01}. The use of galaxies from such 
surveys as tracers at significantly lower redshifts than the CMB have helped to precisely pin down our
cosmological world model \citep[e.g.][]{Per02,Teg04a,Teg04b,Hog05,San06,Teg06}. In particular, such surveys have been used to 
measure the influence of
baryons on galaxy clustering \citep{Col05} and to confirm the potential use of baryon-driven fluctuations (so-called
Baryon Acoustic Oscillations, or BAOs)
in the galaxy power spectrum as a standard ruler with which to set the cosmological distance scale \citep{Eis05}.

The realization that essentially every galaxy hosts a supermassive
black hole \citep[e.g.][]{Mag98,Ric98,Fer00,Geb00}, and that a quasar is therefore just a phase in the normal cycle of a galaxy, 
prompted the more general use of quasars as cosmological tracers. Recent major redshift surveys have also, therefore,
used quasars to probe large-scale structure, with the 2dF QSO Redshift Survey \citep{Cro04} and the Sloan Digital Sky Survey (SDSS)
quasar surveys \citep[e.g.][]{Ric02,Sch10} being notable examples. Quasar redshift surveys have often operated in tandem
with galaxy surveys, and have highlighted the possibility of using quasars to constrain cosmology at
higher redshifts than would be possible for galaxy samples to similar magnitude limits \citep[e.g.][]{Hoy02,Out03,Out04}.

In addition to probing cosmology, quasar clustering can be used as a tool to constrain the interplay of supermassive black holes,
galaxies, and dark matter halos, and how that interplay evolves with cosmic time. Measuring the bias
of quasars constrains the mass of the dark matter halos that quasars occupy. In turn, measuring the abundance of such halos
compared to the number density of the quasars they host can begin to constrain the duration of the quasar phase. In general, this
has led to a consistent picture where UV/optically luminous quasars are biased by a factor of $b_{\rm Q} \sim 2$ at redshift $z\sim1$ rising to 
$b_{\rm Q} \sim 3$ at $z\sim2$ and $b_{\rm Q} > 4$ at $z > 3$
 \citep[e.g.][]{Cro05,Mye07a,Mye07b,Coi07,She07,Ros09,She09,Whi12,ef15}. This implies that UV/optically luminous quasars at $z < 2.5$ are hosted by
halos with an average mass of a few times $10^{12}\,h^{-1} M_{\odot}$ and are ``on'' for a few~per~cent of
the Hubble time. Precise bias and mass measurements for UV/optically luminous quasars at multiple redshifts are crucial in helping to tie down the role of quasars in
galaxy evolution \citep[see, e.g.,][for an overview]{Con13}. In particular, by comparing such clustering measurements to quasar and star-formation 
signatures across the electromagnetic spectrum \citep[e.g.][]{Kru12,Kru15,All14a,All14b,Geo14,Hic14,Dip14,Dip15,DiP16a,Men16}.

As clustering measurements have become increasingly precise, they have become dominated by systematics. Some systematics arise 
from per-cent-level imperfections in calibrating the target imaging or survey spectroscopy that are critical to assembling large redshift catalogs. Other common
systematics include contamination by non-cosmological sources such as stars, or general foregrounds such as Galactic dust. Such systematics
can be scale-dependent, affecting not just the amplitude of clustering measurements, but also the overall shape of the power spectrum of tracers. 
Obviously, this can be a concern for both cosmological constraints and for characterizing the dark matter halos occupied by tracer populations. 
To counter this, procedures have been developed to calculate weighting maps and exclusion masks to ameliorate clustering systematics both for 
galaxies \citep[e.g][]{Ros11,Ros12,Ros16} and for quasars \citep{Mye06,Lei13,Lei14,Aga14,Dip16b,Lau16}. 

The Baryon Oscillation Spectroscopic Survey \citep[BOSS;][]{Daw13} conducted as part of the 
third iteration of the SDSS \citep{Eis11} focused on using quasars and galaxies as complementary probes
of a BAO feature at $\sim100\,h^{-1}$\,Mpc in order to calibrate the redshift-distance relation. 
At redshifts of $z < 0.7$, galaxies were used as direct tracers of the matter power spectrum 
\citep[e.g][]{And12,And14a,And14b} and at $z > 2.1$ clouds of neutral hydrogen in the Lyman-$\alpha$ Forest, as illuminated 
by background quasar-light, were similarly used \citep{Slo11,Slo13,Bus13,Fon14,Del15}. Beyond its cosmological impact, BOSS
provided by far the most precise constraints on the bias and host-halo-masses of quasars at $z\sim2.5$ \citep{Whi12,ef15}.
The success of BOSS led to the development of an extended spectroscopic redshift survey using the SDSS
telescope \citep[extended-BOSS or eBOSS;][]{Dawson+2016}. The cosmological goal
of eBOSS \citep{Zha16} is to detect the $\sim100\,h\,^{-1}$\,Mpc BAO scale in redshift ranges not yet probed by spectroscopic surveys; 
LRGs at $0.7 < z < 0.9$
\citep{Pra16}; Emission Line Galaxies at $z\sim0.9$ \citep{Com15,Rai16} and quasars at $0.9 < z < 3.5$ \citep{Mye15}. In addition
eBOSS will attempt to improve BAO constraints by identifying new quasars to trace the Lyman-$\alpha$ Forest \citep{Pal16}.

Ultimately, eBOSS will provide over half-a-million spectroscopically confirmed
quasars at redshifts of $z > 0.9$ \citep{Mye15}. This sample will provide an unparallelled opportunity to study galaxy evolution and the BAO
scale through quasar clustering, particularly with careful control of the systematics that can contaminate clustering measurements.
In this paper, we present measurements of quasar clustering using the first year of eBOSS observations. The sample that we analyze
approaches 70{,}000 optically luminous quasars in the redshift range $0.9 < z < 2.2$. Even after only its first year, eBOSS has spectroscopically
confirmed $\sim$2--3 times as many $0.9 < z < 2.2$ quasars as used in the main clustering analyses of the 2dF QSO Redshift Survey and the SDSS-I/II
\citep[e.g.][]{Cro05,Ros09,She09}. In this paper, we focus on correcting for the systematics and inhomogeneities that can contaminate eBOSS clustering measurements. We
measure the evolution of quasar bias with unprecedented precision. We then interpret our bias measurements in terms of the characteristic masses
of quasar-hosting halos, and estimate the duty cycle of quasars at $0.9 < z < 2.2$. A companion paper \citep{Rod16arxiv} reexamines our analyses in the context of 
sophisticated N-body simulations.

\section{Data sample}
\label{sec:data}
\subsection{eBOSS survey}  

The six years of observations of eBOSS~\cite{Dawson+2016} started in July 2014. At the end of the survey a sample of more than 500,000 spectroscopically confirmed quasars will be available over 7500 deg$^2$ in the redshift range $0.9<z<2.2$. This will allow for a measurement of the BAO scale and provide measurements of the angular diameter distance, $d_A(z)$, and of $H(z)$ to a 2.8\% and a 4.2\% accuracy, respectively~\cite{Zha16}.
The program also includes 250,000 new luminous red galaxies (LRG) at $\langle z \rangle =0.72$, to be combined with BOSS LRGs and
195,000 emission-line galaxies at $\langle z \rangle =0.87$. Finally the spectra of 60,000 new quasars at $z>2.1$ will be measured and the spectra of 60,000 known quasars at $z>2.1$  will be remeasured to improve their signal-to-noise ratio. This will improve BOSS Lyman-$\alpha$ BAO measurement.

The program makes use of upgraded versions of the SDSS spectrographs \citep{bossspectrometer} mounted on the Sloan 2.5-meter telescope \citep{gunn06} at Apache Point Observatory, New Mexico.
An aluminum plate is set at the focal plane of the telescope with a $3^{\circ}$ diameter field-of-view. Holes are drilled in the plate, corresponding  to 1000 targets, i.e., objects to be observed with one of the two spectrographs. An optical fiber is plugged to each hole and links to the spectrographs. The minimum distance between two fibers on the same plate corresponds to 62'' on the sky, which results in some ``collisions'' between targets. It may, however, be possible to observe both colliding targets if they are in the overlap region between two or more plates.

\subsection{Quasar selection}
\label{sec:TS}

The eBOSS quasar selection~\cite{Mye15} involves a homogeneous CORE selection that combines an optical selection in ({\it u,g,r,i,z}) bands, using a likelihood-based routine called XDQSOz, with a mid-IR-optical color cut. 
The extreme deconvolution (XD) algorithm\footnote{XD~\cite{Bovy+09} is a method to describe the underlying distribution function of a series of points in parameter space (e.g., quasars in color space) by modeling that distribution as a sum of Gaussians convolved with measurement errors.}
was applied in BOSS to model the distributions of quasars and stars in flux space, and hence to separate quasar targets from stellar contaminants~\citep[XDQSO;][]{Bovy+11}.
In eBOSS we use the XDQSOz extension~\cite{Bovy+12}, which selects quasars in any specified redshift range.
We start from the SDSS photometric images in 5 bands ({\it u,g,r,i,z}) \citep{Fukugita+96}  with updated calibration of SDSS imaging relative to BOSS~\cite{Finkbeiner+16}. 
 We select point sources with deextincted PSF magnitudes $g<22$ or $r<22$ that have an XDQSOz  probability $P(z>0.9)>0.2$. 
This selection includes quasars at $z>2.2$, which are not used for direct quasar clustering but for Lyman-$\alpha$ forest studies. There is another quasar selection in eBOSS dedicated to $z>2.2$ Lyman-$\alpha$ quasars, with an average 20 targets per deg$^2$.   
We do not discuss it here.

In contrast to quasars, stars tends to be dim in the mid-IR wavelengths. We make a weighted average of the WISE~\cite{Wright+10} $w1$ and $w2$ mid-IR fluxes to optimize the $S/N$ ratio and similarly a weighted average of SDSS $g$, $r$ and $i$ PSF fluxes. Selecting targets with a resulting  average optical magnitude significantly larger than the IR magnitude such that
$m_{\rm opt} -m_{\rm IR} \ge (g-i) +3$, reduces the star contamination in our sample without significantly removing quasars.

This selection results in an average 115 targets per deg$^2$, out of which 25 have already been observed by SDSS I, II or III, so there remain 90 targets per deg$^2$ to be measured by eBOSS. 
These 25 and 90 targets per deg$^2$ result, respectively, in 13 quasars per deg$^2$ that we call "known quasars'' and 58 new quasars per deg$^2$ that we call ``eBOSS quasars'', in both case in the range $0.9<z<2.2$.
This makes a total of about 70 quasars per deg$^2$ and matches the requirement to reach a 2\% accuracy on the BAO scale~\cite{Dawson+2016}. 

Part of the eBOSS footprint was observed by SEQUELS~\cite{Dawson+2016,Mye15}, 
a pilot survey at the end of SDSS III. SEQUELS differs from the rest of eBOSS survey in two ways: the apatial placement of the plates was denser, one plate per $\approx 4$ deg$^2$ instead of one plate per $\approx 5$ deg$^2$, and all quasar-target spectra were visually inspected. 
The first difference is taken into account by the completeness (see section~\ref{sec:computing_xi})
and, in order to treat them as all eBOSS quasars,
we use only the pipeline information for SEQUELS quasars. 

The data used in this paper include all spectra taken during the first year of eBOSS data taking, up to July 2015. They cover a surface of 1200 deg$^2$.











\begin{figure}[t]
\begin{center}
\epsfig{figure=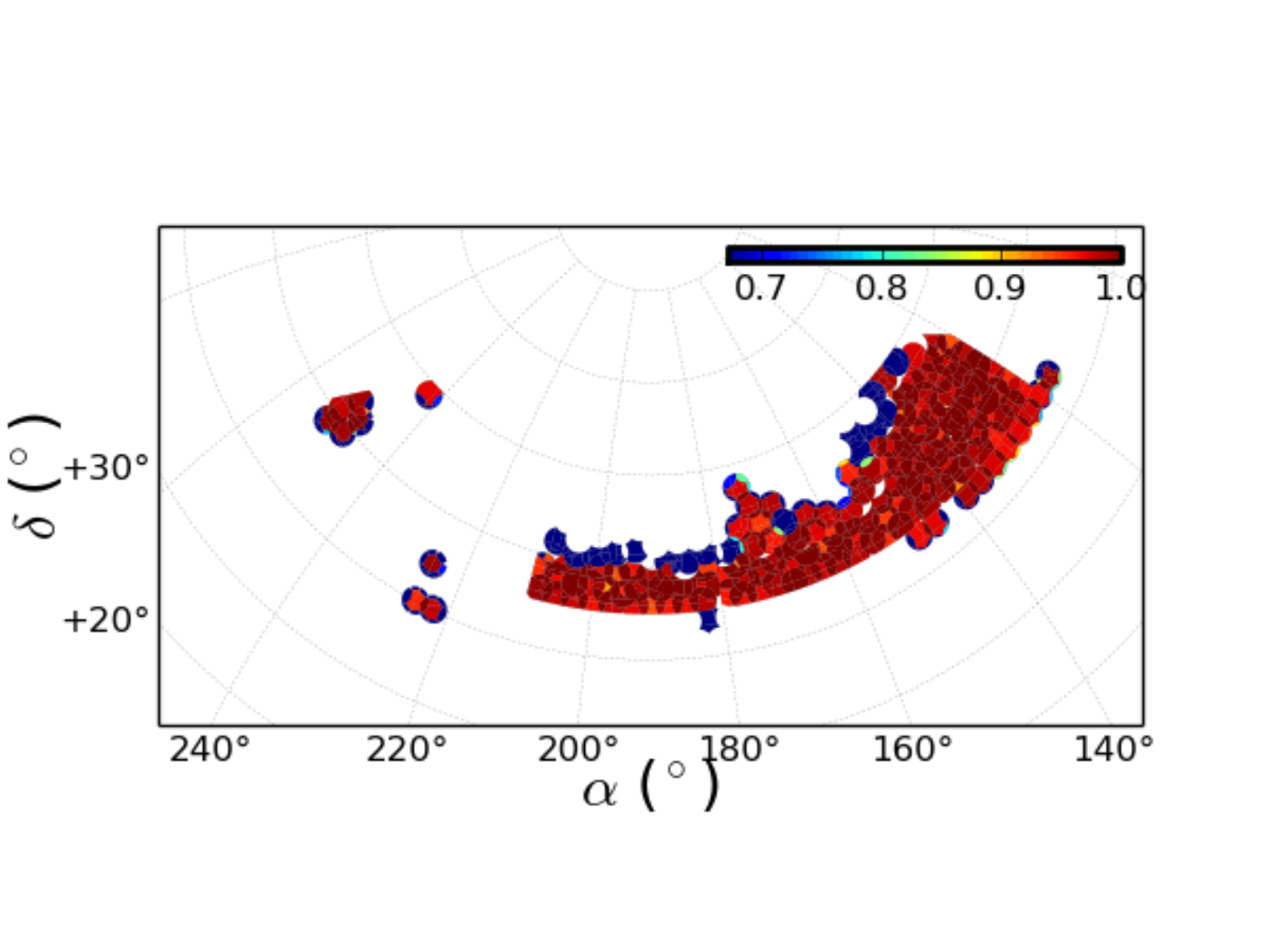 ,width = 10cm} 
\epsfig{figure=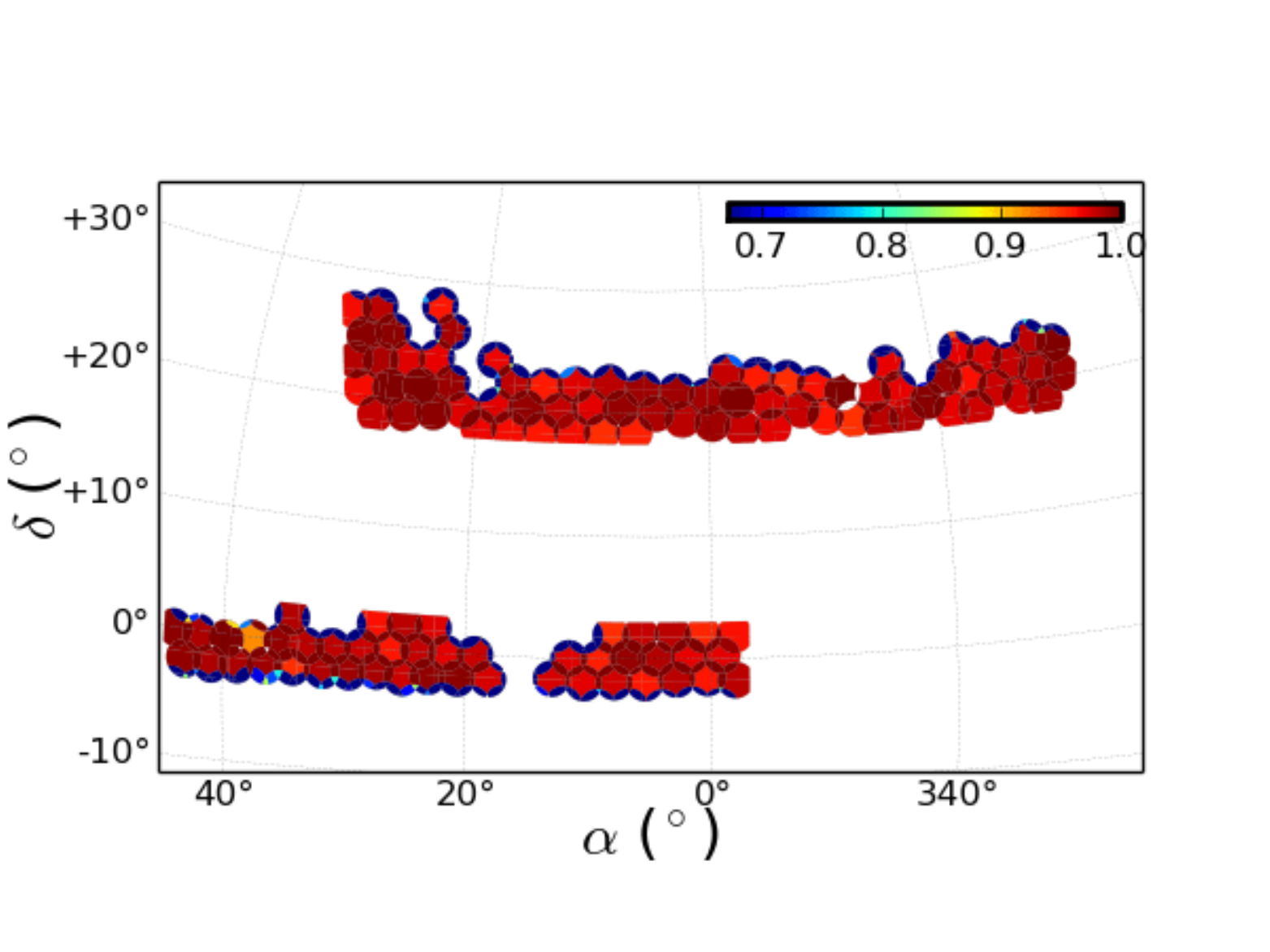 ,width = 10cm} 

\caption{\it  Angular distribution of the selected data in the NGC (left) and the SGC (right). 
The color scale indicates the survey completeness in each polygon. }  
\label{fig:footprint}
\end{center}
\end{figure}

\begin{figure}[t]
\begin{center}
\epsfig{figure=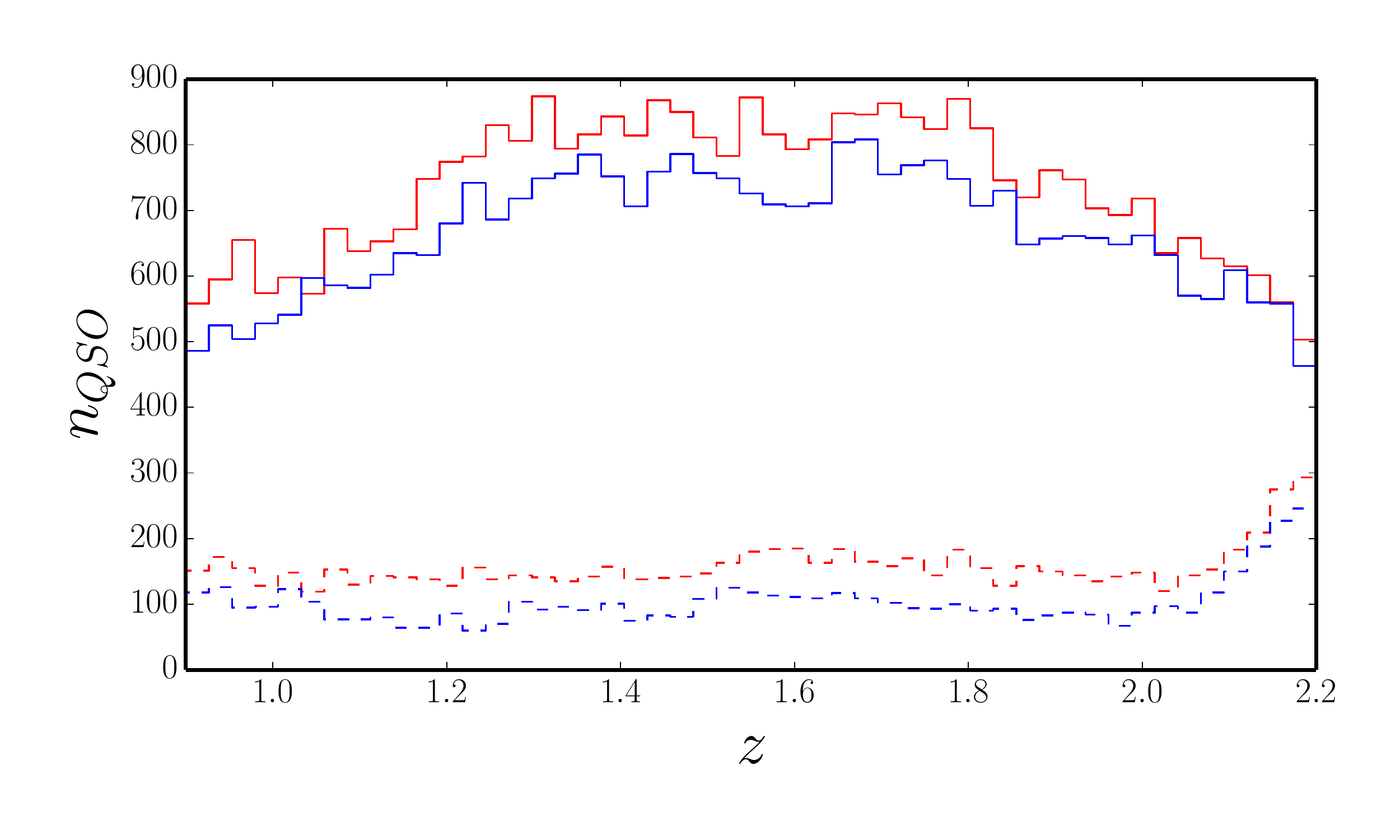 ,width = 8cm} 
\caption{\it  Redshift distribution of eBOSS quasars (continuous line) and known quasars (dashed line) in the NGC (red)  and SGC (blue).}
\label{fig:nz}
\end{center}
\end{figure}

\section{Analysis}
\subsection{Computing $\xi(r)$}
\label{sec:computing_xi}

There are a limited number of fibers available so all targets cannot be ascribed a fiber and observed.
Since the density of eBOSS targets is not homogeneous, their probability to be observed is not homogeneous either.  
In addition targets are more likely to be observed when located on areas where plates overlap. In order to take those effects into account we define polygons as the intersections of the plates projected on the celestial sphere, and in each polygon we define a completeness
\begin{equation}
C = \frac{N_{\rm obs} + N_{\rm col}}{N_{\rm targets} - N_{\rm known}} \,.
\label{eq:comp}
\end{equation}
Here $N_{\rm obs}$ is the number of observed targets, $N_{\rm targets}$ the total number of targets, $N_{\rm known}$ the number of targets that have already been observed by the SDSS I, II and BOSS surveys, and $N_{\rm col}$ is the number of targets that were not observed because they are colliding with a quasar.
Known targets are not re-observed by eBOSS in order to save fibers, and are thus removed from the denominator of equation \ref{eq:comp}. Besides, known target completeness is by definition equal to 1, which would bias our measurement. In order to force the known-target completeness to be the same as for other targets, we remove some known targets from the sample with a survival probability equal to the value of the completeness in their polygon. 
We account for collisions as in \citet{Anderson12} : when a target is not observed due to collision, we upweight by one unit the closest observed quasar within 62'' (for limitations of this approach, see \citet{Bianchi17}). Therefore we add these collided targets to the numerator when computing the completeness. If there is no quasar within 62'', this target is treated as any other unobserved targets. 
%
Figure~\ref{fig:footprint} shows the completeness of the eBOSS survey in the North Galactic Cap (NGC) and South Galactic Cap (SGC), as computed with the Mangle software~\cite{mangle}.

To correct for completeness, we generate a catalog of $10^{7}$ objects with ``random'' angular positions over the eBOSS footprint, with the number of random objects in each polygon proportional to its area times its completeness. We then assign to each random object a redshift that is drawn from the measured redshift distribution $n(z)$, see Figure~\ref{fig:nz}. Finally, we compute $\xi(r)$ with the Landy-Szalay estimator \cite{LandySzalay93} :

\begin{equation}
\hat{\xi}^{\rm LS}(r) = \frac{dd(r) - 2 dr(r) + rr(r)}{rr(r)} \,  ,
\label{eq:LS_estimator}
\end{equation}
where $dd(r)$ is the number of pairs of quasars separated by a distance $r$, $dr(r)$ is the number of pairs between a quasar and an object from the random catalog, and $rr(r)$ the number of pairs of random objects. These three quantities are normalized to the total number of pairs.

As mentioned in section \ref{sec:data}, the measure of the correlation function $\xi(r)$ is very sensitive to inhomogeneities in the quasar target selection. We apply masks to remove from our sample all quasars and random objects that are located in areas where the target selection is too contaminated to be modelled and easily corrected. These areas include regions around bright objects (stars or galaxies) 
and where the SDSS photometry is unreliable. 
We also remove areas covered by the centerpost of the eBOSS plates, since we cannot observe those areas.

\subsection{Estimation of statistical uncertainties}


We compare two methods to compute covariance matrices. The first method, developed by~\citet{Lau16}, uses bootstrap realizations. For each galactic cap, we define 201 bootstrap cells.  We obtain a bootstrap realization by drawing 201 cells with replacement from the 201 bootstrap cells, and compute $\xi(r)$ for this realization. We repeat this operation 10,000 times, and estimate the covariance matrix of $\xi(r)$ from the covariance of $\xi(r)$ for these 10,000 realizations. Bootstrap resampling ignores cosmic variance, but this is not an issue here since our sample is shot-noise limited. Finally, we note that computing the covariance matrix from data resampling means that it includes variations caused by systematic effects present in the data.

We also compute covariance matrices using 100 QPM mocks \cite{QPMmocks} for each galactic cap. These mocks take into account cosmic variance, but they struggle to model the correlation function on small scales. However, this is not problematic because these scales are not relevant for this study. 

\begin{figure}[t]
\begin{center}
\epsfig{figure=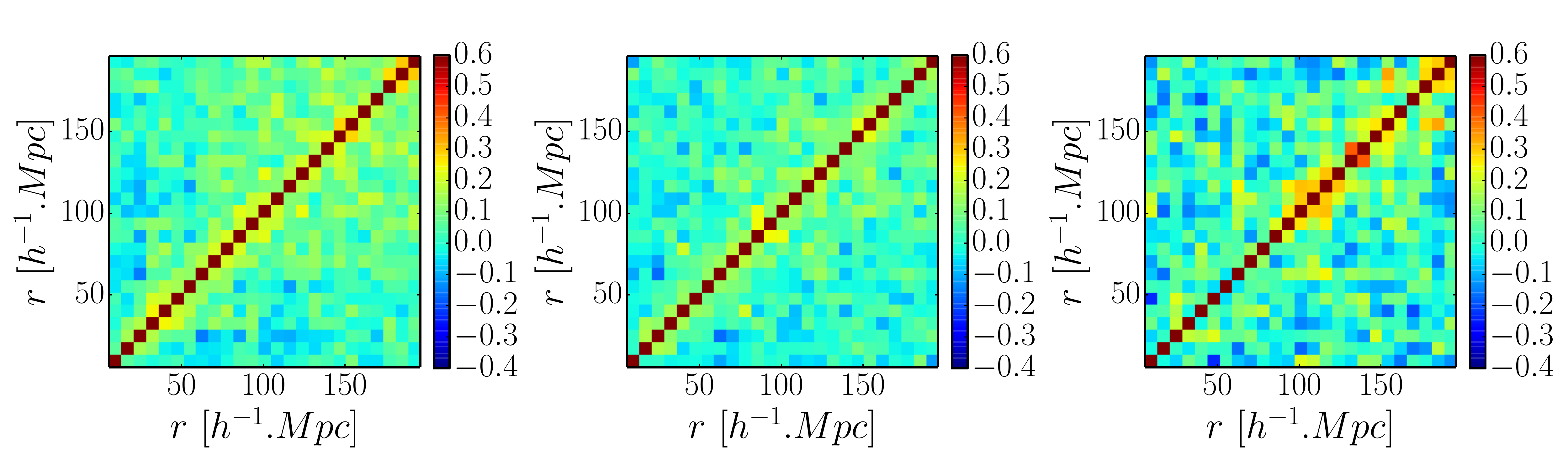 ,width = 15cm} 
\caption{\it Correlation matrices of $\xi(r)$ computed using bootstrap realizations without systematic weighting (left), with systematic weighting (center), and using QPM mocks (right).}
\label{fig:correlation}
\end{center}
\end{figure}

\begin{figure}[t]
\begin{center}
\epsfig{figure=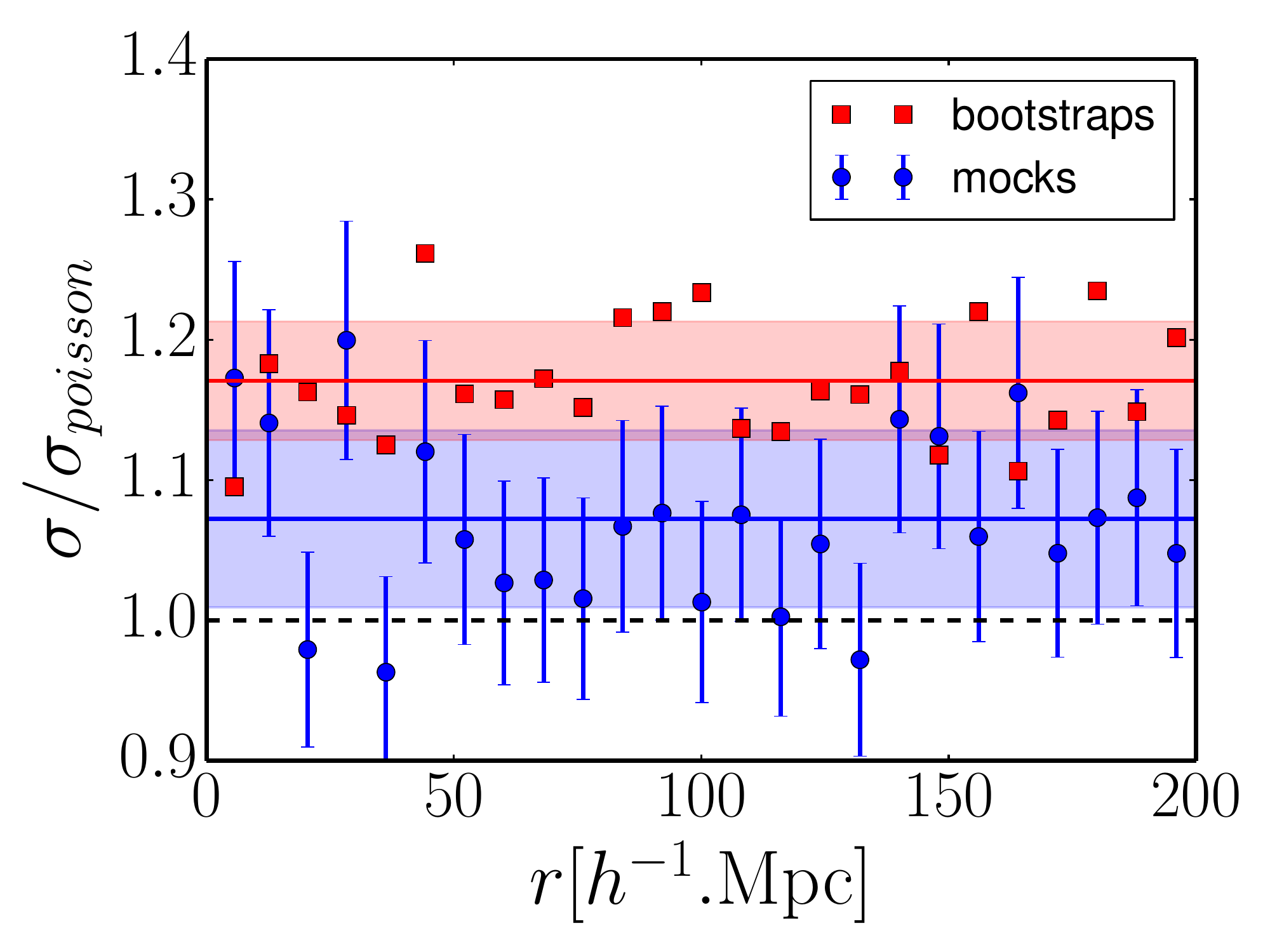 ,width = 8cm} 
\caption{\it Ratio of bootstrap errors to Poisson errors (red squares) and, ratio of mock errors to Poisson errors (blue dots). The lines correspond to the mean value of the ratio, and the bands represent the rms of the dots.}
\label{fig:comp_poisson}
\end{center}
\end{figure}

Figure \ref{fig:correlation} displays the correlation matrices of $\xi(r)$ for the full eBOSS survey obtained with the mocks and the bootstrap realizations. We note that systematic weighting (Section~\ref{sec:Systematics}) slightly reduces the amplitude of the off-diagonal elements of the bootstrap correlation matrices on large scales (center compared to left). The mock correlation matrix is noisier because we only have 100 mock catalogues.
Figure \ref{fig:comp_poisson} shows the ratio of bootstrap and mock errors to Poisson errors. We see that bootstrap errors are systematically larger than mock errors, but provide a more accurate determination of uncertainties. In the following, we will always display statistical uncertainties obtained from bootstrap realizations.



\section{Systematic effects}
\label{sec:Systematics}

\subsection{Inhomogeneities of target identification}
The SDSS I-II and BOSS surveys observed a total of 12,759 quasars within our target sample.
These ``known quasars'' were spectroscopically identified by visual inspection \cite{Par16}, whereas newly observed targets are identified automatically by the eBOSS pipeline. The efficiency of target identification is known to be better for the visual inspection than for the pipeline, and such a difference can generate systematic effects. This efficiency also depends on the signal-to-noise ratio, which varies with the position of the fiber in the spectrographs. Fibers with an identifier, $n_{\rm ID}$, close to 0, 500 or 1000 are located on the edges of the spectrographs, and their spectra are on average noisier than for other fibers. Since the fiber IDs are also correlated with the position of the fiber in the focal plane, the difference in noise can generate correlations at scales of the order of the plate width.

\begin{figure}[t]
\begin{center}
\epsfig{figure=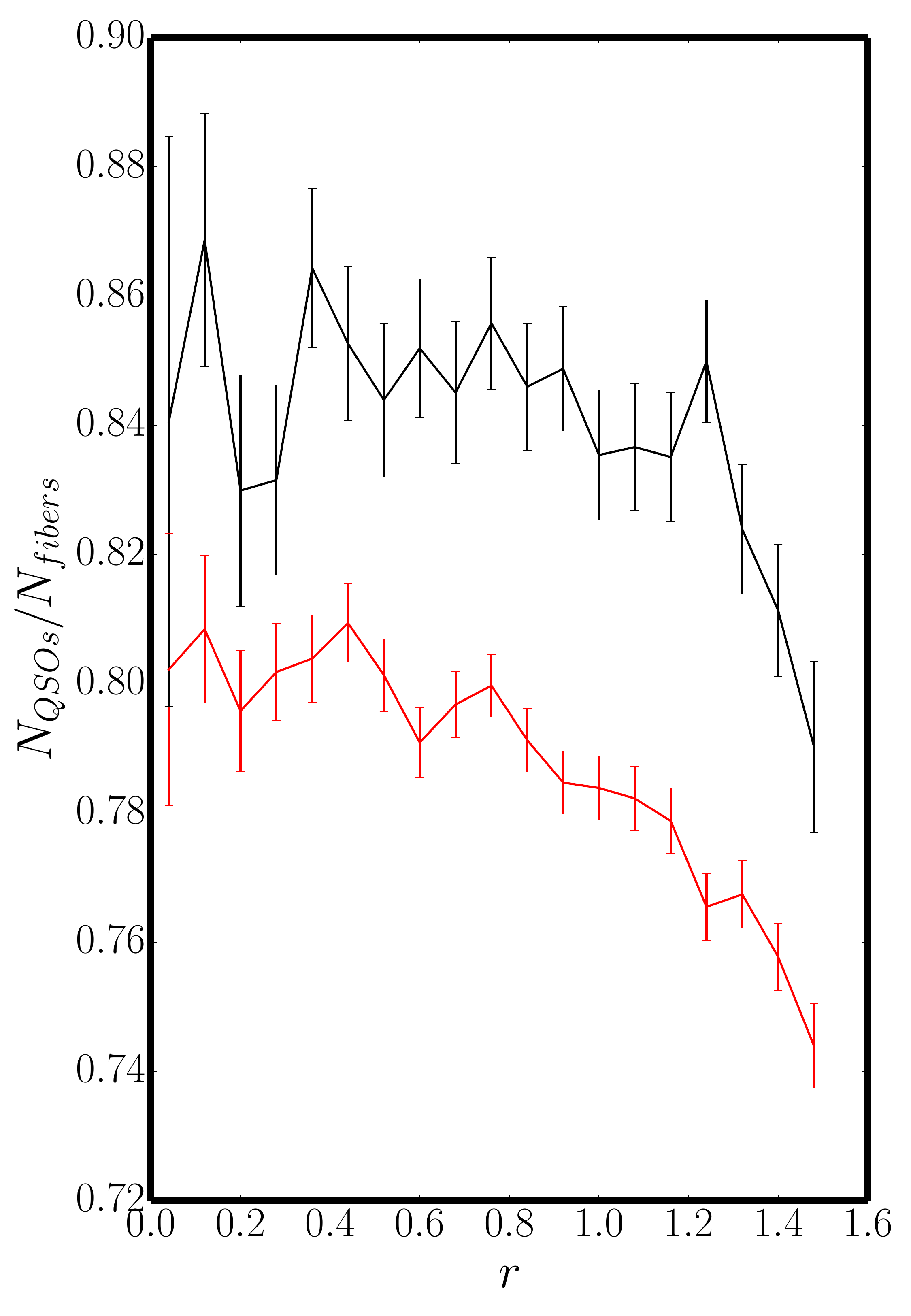,width = 5cm}
\epsfig{figure=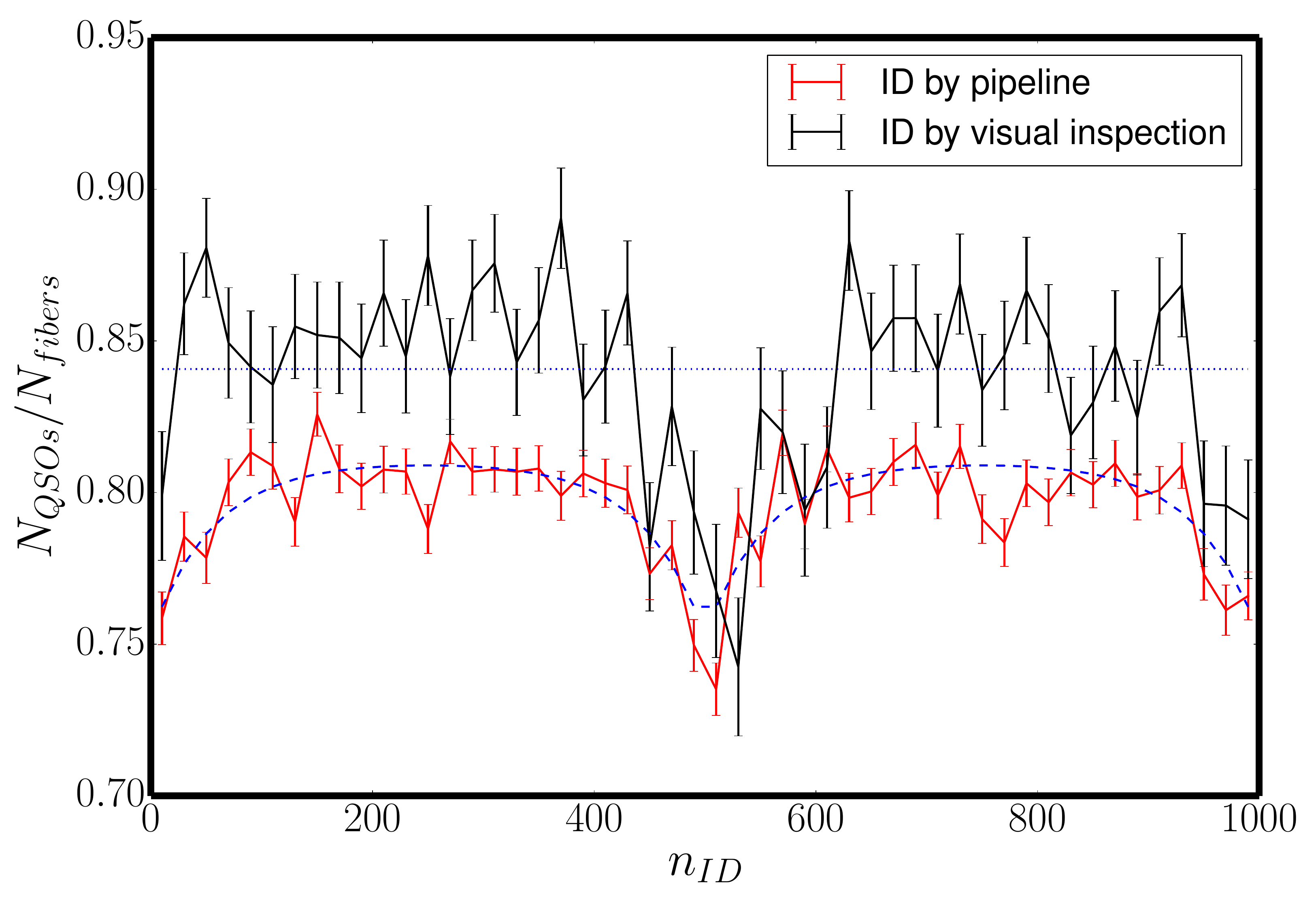,width = 7cm}
\caption{\it The $N_{QSO} / N_{fibers}$ ratio with respect to $r$ (left, in radians) and $n_{\rm ID}$ (right). The black lines correspond to the visually inspected quasar sample, the red lines correspond to the new eBOSS quasar sample, identified by the pipeline. The dashed blue line corresponds to a fit of the function in equation~\ref{eq:fiberid} on the new eBOSS quasar sample, and the dotted blue lines corresponds to a fit of a constant on the visually inspected quasar sample.} 
\label{fig:syst_fiberID_r}
\end{center}
\end{figure}

Figure~\ref{fig:syst_fiberID_r} shows the ratio of the number of identified quasars to the number of fibers, $N_{\rm QSO}/N_{\rm fibers}$, versus the distance to the center of the plate, $r$, and versus the fiber ID, $N_{\rm ID}$. The red lines correspond to the newly observed eBOSS targets, identified by the pipeline. The black lines correspond to the targets that have been visually inspected, including all known eBOSS quasars.

 As expected, the ratio $N_{\rm QSO}/N_{\rm fibers}$ is higher for visually inspected targets than for targets identified with the pipeline, and both ratios present a significant variation with respect to $n_{\rm ID}$ and $r$. The dependency with $n_{\rm ID}$ is well fitted by a hyperbolic cosine with 3 free parameters: 

\begin{equation}
\frac{N_{ \rm QSO}}{N_{ \rm fibers}}(n_{\rm ID}) = -a \cosh \left( \frac{n_{\rm ID} \, {\rm mod} \, 500 - 250}{b}\right)^{2} + c \,.
\label{eq:fiberid}
\end{equation}

The resulting fit corresponds to the dashed blue line of Figure~\ref{fig:syst_fiberID_r}. We correct this effect by weighting eBOSS quasars by the inverse of equation \ref{eq:fiberid}. Since we do not know the fiber number for all known quasars, we simply weight them with the inverse of the mean value of the ratio, displayed by the blue dotted straight line 
 on Figure~\ref{fig:syst_fiberID_r}. This takes into account the difference of efficiency of identification between known and new eBOSS quasars. Comparing blue points with magenta dots in Figure~\ref{fig:xi_weighting} shows that this unidentification weighting has little effect on $\xi(r)$.
In addition we also tried a weighting scheme where the hyperbolic cosine dependency for new eBOSS quasars is replaced by the mean value of the ratio, as is done for known quasars. This showed that most of the (small) effect of the unidentification weight comes from the difference of efficiency between the two categories of targets rather than from the dependence of the weight with $n_{\rm ID}$. So neglecting the dependence of the weight with $n_{\rm ID}$ for known quasars is safe.




\subsection{Inhomogeneities of quasar target selection}

\begin{figure}[t]
\begin{center}
\epsfig{figure=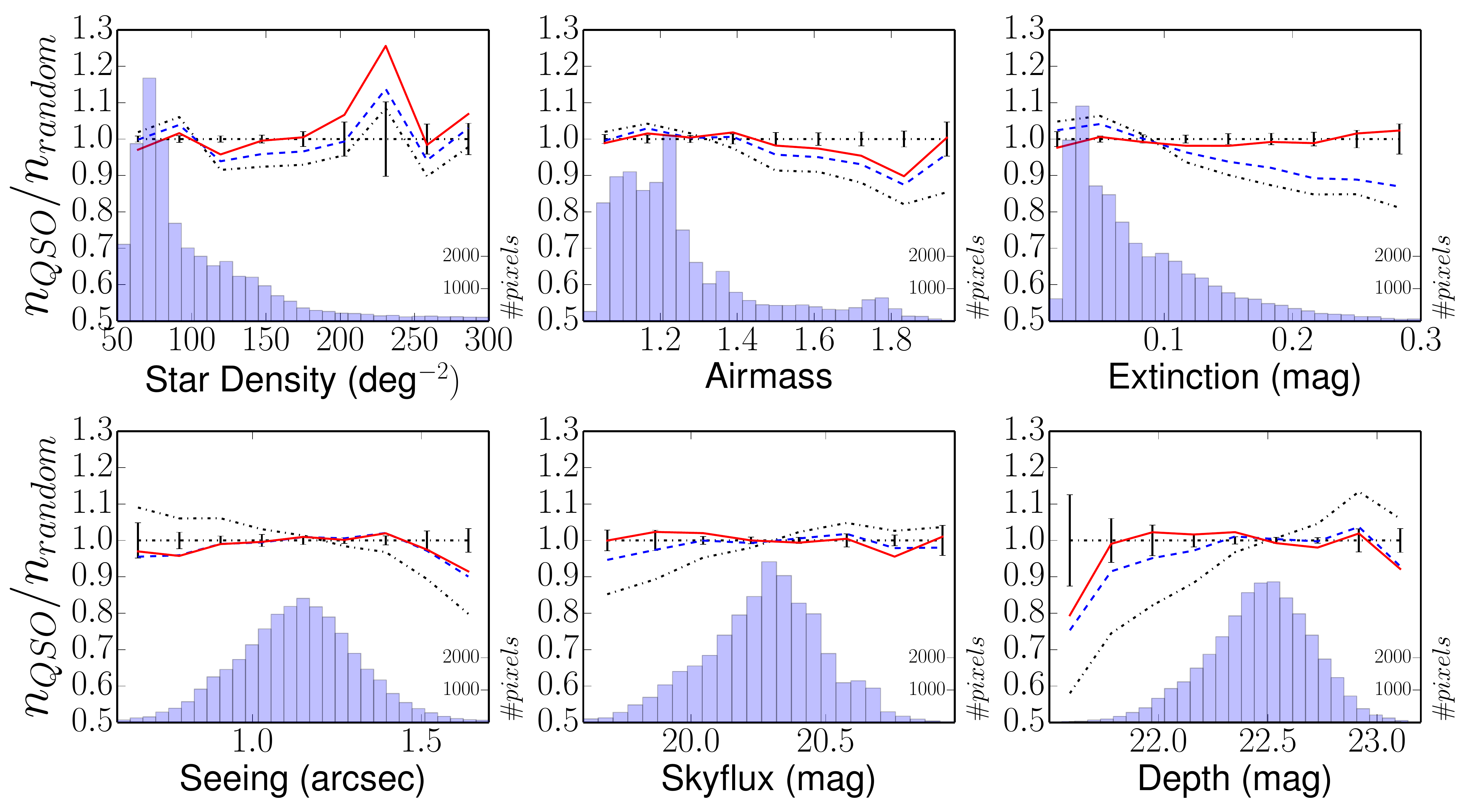,width = 13cm} 
\caption{\it The $n_{\rm QSO}/n_{\rm randoms}$ ratio versus star density, airmass, Galactic extinction, seeing, sky-flux and depth, for the i-band before any correction (black dotted lines), after correction by the depth (blue dotted lines), and after correction by the depth and the extinction (red full lines). For the sake of clarity, the error bars, which are similar before and after weighting, are only displayed once on the $n_{\rm QSO}/n_{\rm randoms} = 1$ line. The histograms display the distribution of pixels for the healpix maps.}
\label{fig:syst_dependencies}
\end{center}
\end{figure}

Quasar targets are selected with the XDQSOz algorithm (section~\ref{sec:TS}), which aims at providing a homogeneous target selection using the SDSS photometry. The SDSS photometry, however, is not perfectly homogeneous. The mean $5 \sigma$ detection limit for a point source (also called depth) for the SDSS photometry is $g = 23.1$ and $r = 22.7$, but it varies with angular position by up to $\pm 0.8$ magnitude (see histogram on Figure~\ref{fig:syst_dependencies}). Targets are selected up to a given apparent magnitude limit of $g = 22.0$ or $r = 22.0$, therefore some faint sources can end up very close to the detection limit. Uncertainties on their relative flux measurements will be significantly higher than for other sources, so the XDQSOz probability of faint sources may go below the selection threshold. Also, observed fluxes of faint sources might be biased by the fluxes of close brighter sources, an effect known as blending.

We study the variation of the observed-quasar density with the depth and its inputs (seeing, airmass, Galactic extinction and sky-flux). We also study the variation with star density, since it can bias the target selection through blending. To do so, we generate Healpix maps with $N_{\rm side} = 256$ for each of the aforementioned quantities, following the procedure of \textit{Ross et al. 2012} \cite{Ros12}. We also create a map for the ratio of the number of observed quasars to the normalized number of random objects : this quantity is proportional to the observed-quasar density corrected for completeness. The black dotted lines on  Figure~\ref{fig:syst_dependencies} show that this ratio varies with all quantities, except star density. This means that we do not observe any bias in the quasar target selection due to blending effects. The dependencies are compatible between the NGC and the SGC, and they do not depend on redshift.


We fit a linear function to the dependency with the depth. We weight each quasar with the inverse of the fitted function for the value of the depth in the considered pixel of the map, and recompute the $n_{\rm QSO} / n_{\rm random}$ ratio. The blue lines in Figure~\ref{fig:syst_dependencies} show that the dependencies of this ratio with airmass, seeing, sky-flux and depth vanish, and that the dependency with Galactic extinction is reduced, but still significant. The same procedure is applied to the observed-quasar density already corrected for depth to correct for this remaining dependency with Galactic extinction.
The final systematic weights for target selection inhomogeneity are obtained by multiplying the depth and Galactic-extinction weights.

\subsection{Effect of weightings on $\xi(r)$}

\begin{figure}[t]
\begin{center}
\epsfig{figure=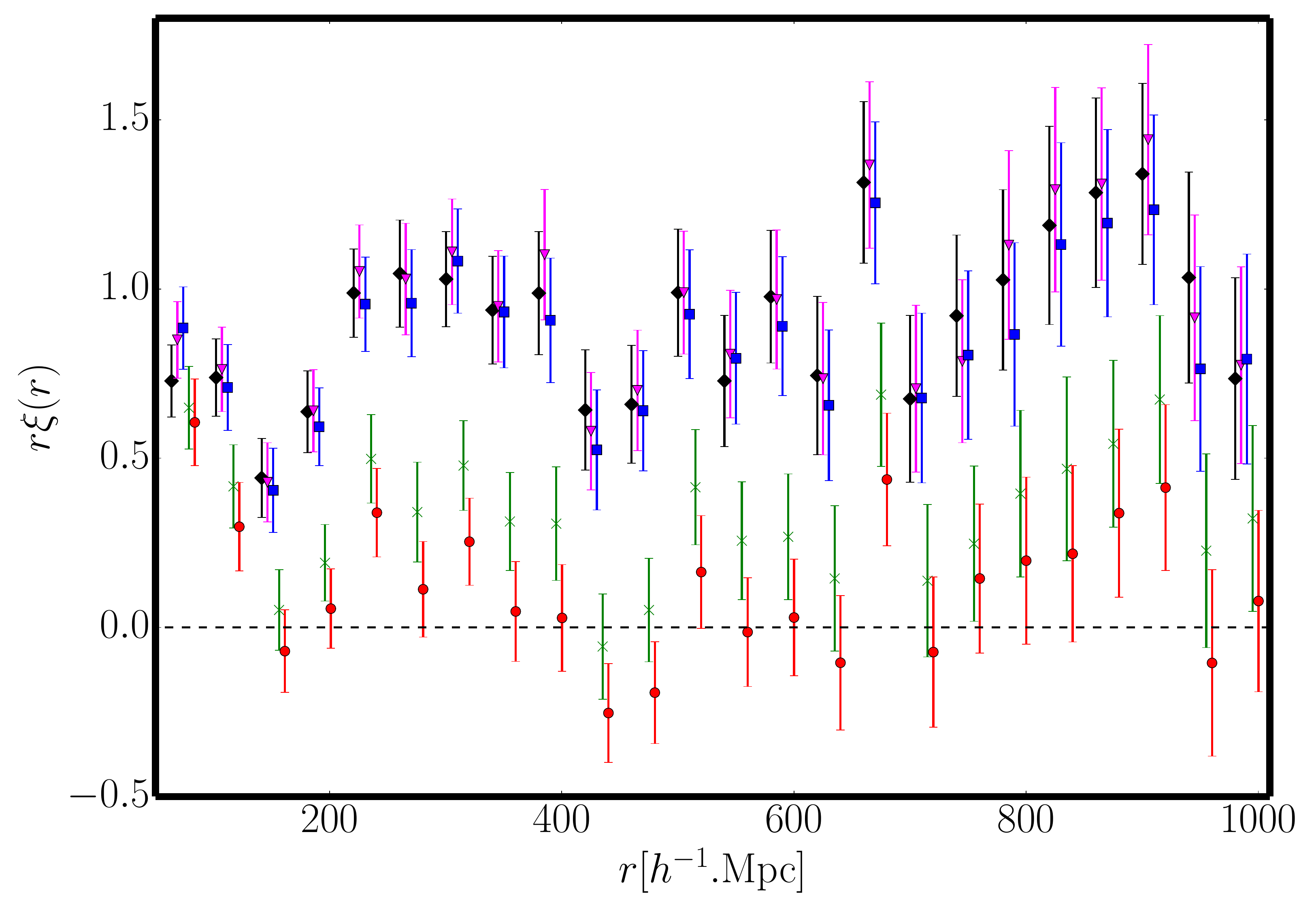,width = 10cm} 
\caption{\it Effects of the different weightings on the quasar correlation function. The black lozenges correspond to $\xi(r)$ without any weighting, the magenta triangles to $\xi(r)$ with the collision weights. The blue squares are obtained with the combination of the collision and unidentification weights, the green crosses are obtained with the addition of the depth weights and the red circles are obtained with the addition of the depth and Galactic-extinction weights.}
\label{fig:xi_weighting}
\end{center}
\end{figure}

Figure \ref{fig:xi_weighting} shows $r \cdot \xi(r)$ without any weights, and with the successive addition of collision weights, unidentification weights, depth weights and depth plus Galactic-extinction weights. We quantify the effect of the corrections by computing the cross-$\chi^{2}$ between $\xi_{N}(r)$ and $\xi_{S}(r)$, the correlation functions measured in the NGC and the SGC :

\begin{equation}
\chi^{2}_{\rm NS} = \sum_{ij}(\xi_{\rm N}(r_{i}) - \xi_{\rm S}(r_{i})) C^{-1}_{ij} (\xi_{\rm N}(r_{j}) - \xi_{\rm S}(r_{j})) \, ,
\label{eq:cross_chi2}
\end{equation}
where $C$ is the sum of $C_{N}$ et $C_{S}$, the covariance matrices of $\xi_{N}(r)$ and $\xi_{S}(r)$. The resulting values are shown in Table~\ref{tab:cross_chi}. The main effect clearly arises from weighting with the depth, which strongly reduces the value of $\chi^{2}_{\rm NS}$. The correction for fiber collisions has only a limited impact on larger scales, and is not susceptible to bias the measure of $b_{Q}$. In the following, we will always apply the full weighting scheme to our data sample.

\begin{table}
\centering
\begin{tabular}{l|c} \hline
Weighting scheme & $\chi^{2}_{\rm NS} (24 \, \rm d.o.f.)$ \\
\hline
No weights & 161 \\
Collision & 154 \\
Collision + Unidentification & 128\\
Collision + Unidentification + Depth & 58 \\
Collision + Unidentification + Depth + Galactic extinction & 47 \\

\hline
\end{tabular}
\caption{Values of $\chi^{2}_{\rm NS}$ (see Eq.~\ref{eq:cross_chi2}) for different weighting schemes.}
\label{tab:cross_chi}
\end{table}

\section{Measurement of the quasar bias}

In order to measure the quasar bias, $b_{Q}$, we fit the measured $\xi(r)$ with a flat $\Lambda$CDM model, using the same cosmological parameters used for the BOSS twelfth Data Release (namely $h=0.676$, $\Omega_m=0.31$, $\Omega_b h^2=0.0220$, $n_s=0.9619$). 
Using these parameters, CAMB \cite{CAMB} and HALOFIT \cite{HALOFIT} provide a non-linear matter power spectrum $P_{mat}(k)$. We account for linear redshift-space-distortions using the Kaiser formula \cite{Kaiser87} :

\begin{equation}
\label{Eq:Pmodel}
P_{Q}(k,\mu)=b_{Q}^2\; (1+\beta\mu_{k}^2)^2\;P_{mat}(k) \;,
\end{equation}
where $\mu_k$ is the cosine of the angle between $k$ and the line of sight, $\beta = f/b_{Q}$, and $f \simeq \Omega_{m}^{0.55}(z)$ is the growth rate of structures. The last step consists in converting $P_{Q}(k)$ into $\xi_{Q}(r)$ using a Fast Fourier Transform (FFT).

\begin{figure}[t]
\begin{center}

\epsfig{figure=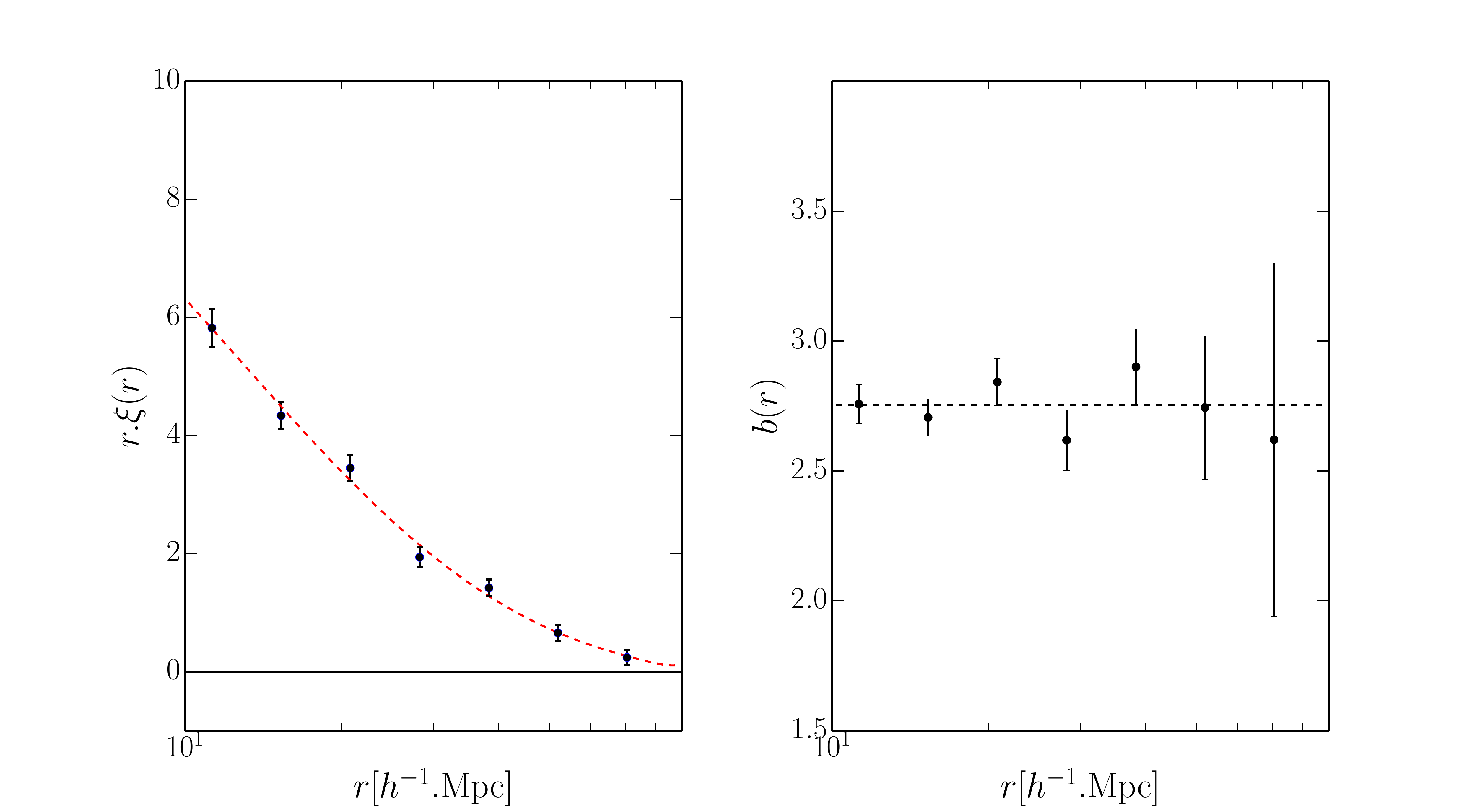,width = 15cm} 
\caption{\it Left : $\xi(r)$ for the eBOSS survey (black dots). The red dashed curve is a $\Lambda$CDM fit to the data. Right~:~measured quasar effective bias ($\xi_{QSO} / \xi_{mat, \rm CAMB}$) as a function of the separation $r$. The bias is compatible with a constant, even on larger scales due to systematic weighting.} 
\label{fig:quasar_bias}
\end{center}
\end{figure}

All fits of $b_{Q}$ are performed using the MINUIT libraries \cite{MINUIT} over the range $10<r<85$ $h^{-1}$Mpc. The fit, shown on Figure~\ref{fig:quasar_bias}, exhibits a fair agreement with the $\Lambda$CDM model ($\chi^{2} = 4.0$ for 7 d.o.f.). For the full eBOSS survey, we measure $b_{Q} = 2.45 \pm 0.05$, for $\bar{z} = 1.55$. This result is in agreement with the results obtained by~\citet{Cro05} using the 2dF QSO Redshift Survey : their empirical parametrization yields a value $b_{Q}(z=1.55) = 2.41$. \citet{Cro05} give the error on the two parameters of their fit but not the correlation. If we neglect the correlation, the error on $b_{Q}(z=1.55)$ is 0.30. In any case our measurement is compatible with their parametrization. If we fit their data, we confirm their values of $a$ and $b$, and find a correlation coefficient $\rho_{a,b} = -0.90$. Taking into account this anticorrelation, yields a much lower error on $b_Q(z=1.55)$ of 0.10. In any case our measurement is compatible with their parametrization.

These results are also compatible with the measurement obtained with the SDSS II quasar sample. The right panel of Figure~\ref{fig:quasar_bias} shows that the ratio $\xi_{QSO} / \xi_{mat} = b_{Q}^{2}(1 + \frac{2}{3} \beta + \frac{1}{5} \beta^{2})$, where $\xi_{mat}$ is the matter correlation function, remains nearly constant with $r$. This means that our measurement of $b_{Q}$ is not sensitive to the range of the fit.



\begin{figure}[t]
\begin{center}
\epsfig{figure=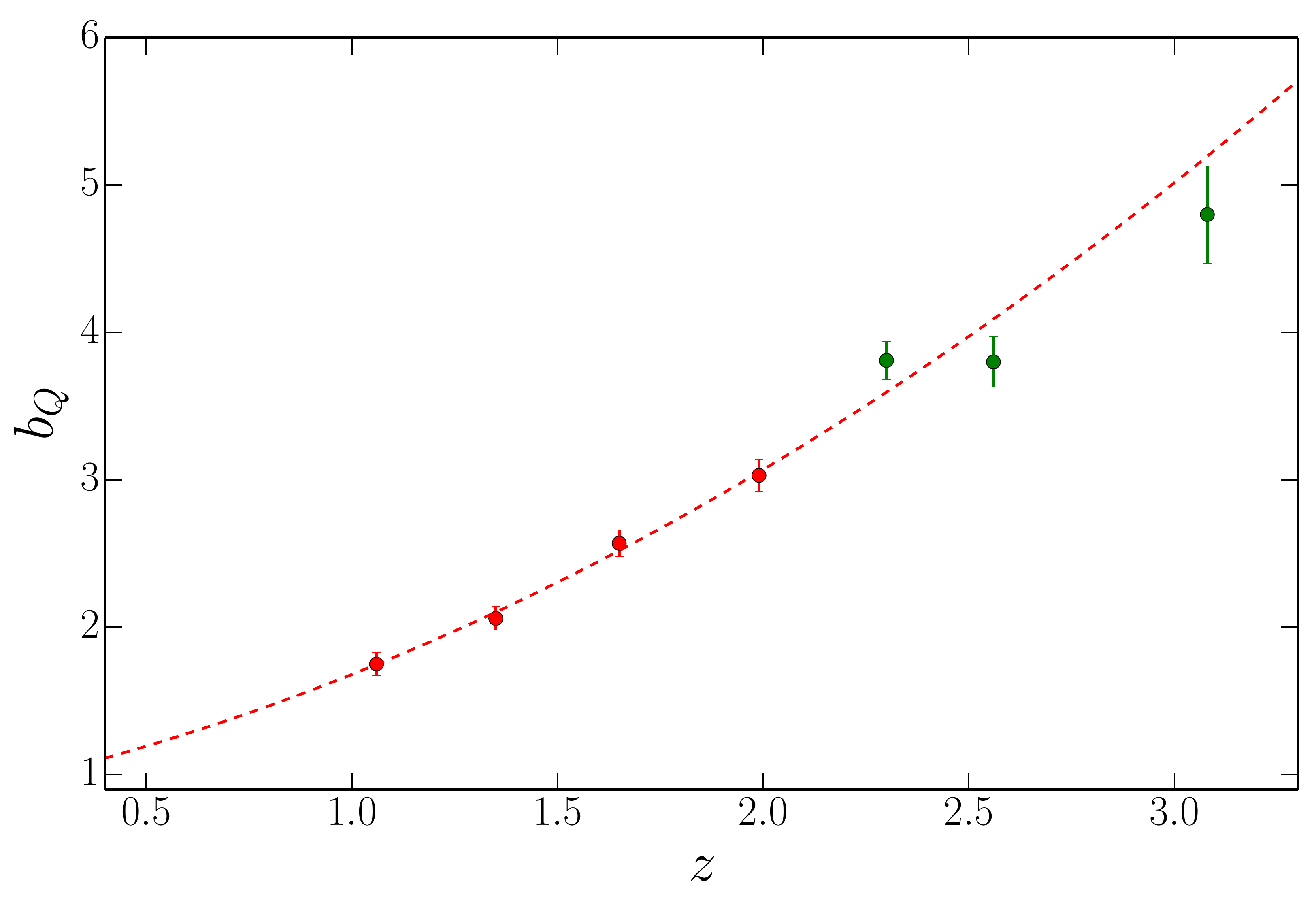,width = 6cm} 
\epsfig{figure=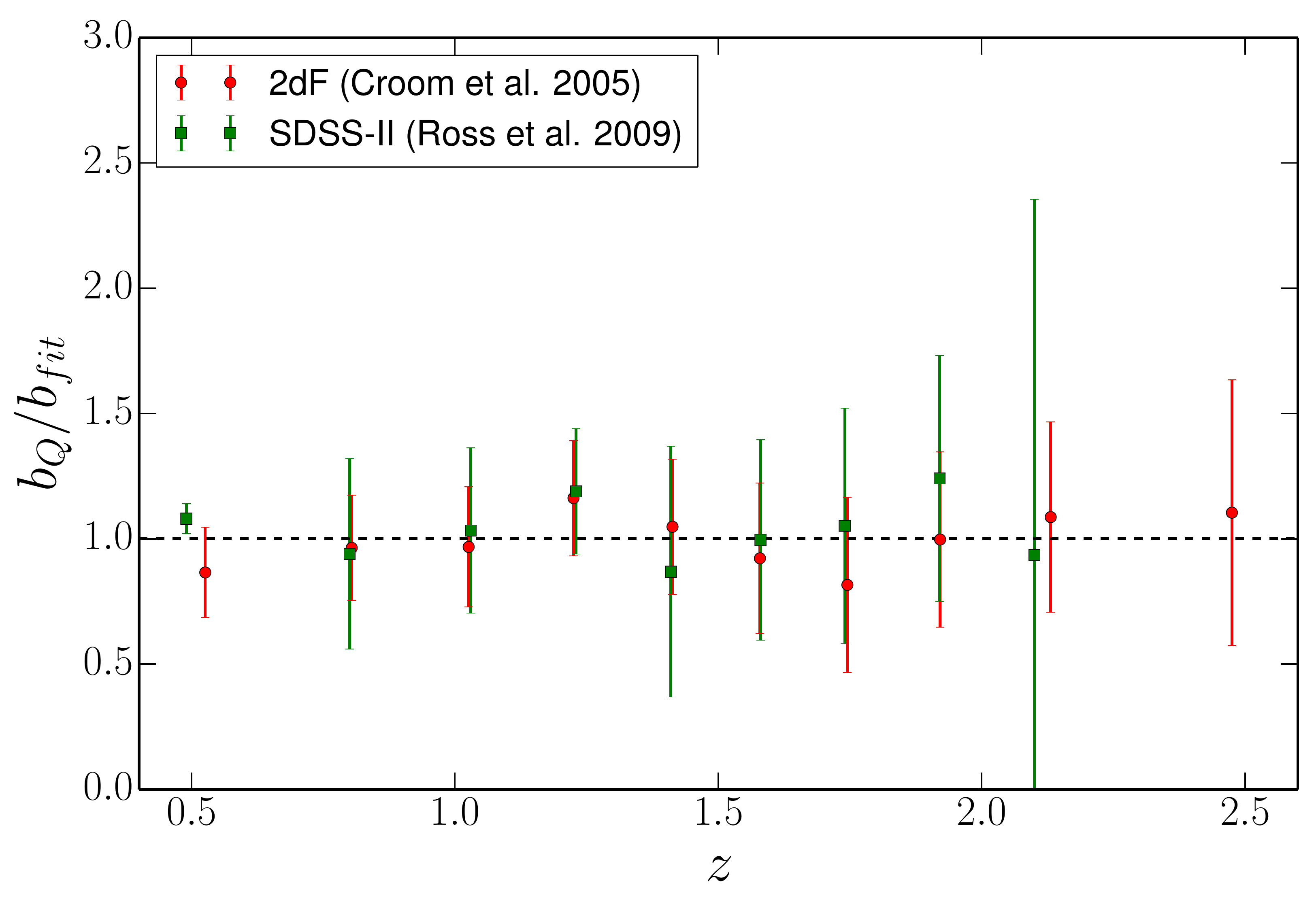,width = 6cm}
\caption{\it Left : measured quasar bias as a function of redshift. The red dots correspond to this analysis (eBOSS data) and the green dots correspond to the bias measured with the BOSS sample, in the range $2.2 < z < 3.5$~\cite{Lau16}. The red dotted line is the fit $b_{Q}(z) = a(1+z)^{2}+b$. Right : ratio of quasar bias measured by previous surveys to the value of the function fitted on the results of the eBOSS and BOSS samples. The red dots correspond to the 2dF sample, the blue squares to the SDSS-II quasar sample.} 
\label{fig:b_z}
\end{center}
\end{figure}

We cut our sample in 4 redshift slices, and measure $b_{Q}$ in each subsample : the results are displayed on Figure~\ref{fig:b_z}, alongside results from the BOSS quasar sample. The numerical values are presented in Table~\ref{tab:b_z}. We combine the measurements of $b_{Q}$ from the eBOSS and BOSS samples, and fit $b_{Q}(z)$. In order to avoid a large anti-correlation between the fit parameters obtained with Croom parametrization, we use an equivalent parametrization defined such as to yield non correlated parameters :

\begin{equation}
b_Q(z) = \alpha [(1+z)^2 - 6.565)] + \beta \\
\end{equation}

with 

\begin{equation}
\alpha=0.278 \pm 0.018, \quad \beta=2.393 \pm 0.042, \quad \rho_{\alpha,\beta}=0 \, ,
\end{equation}

where $\rho_{\alpha,\beta}$ is the correlation coefficient between the parameters $\alpha$ and $\beta$. This is equivalent to $a = 0.278 \pm 0.018$, $b =  0.57  \pm 0.13$ and $\rho_{a,b}=-0.94$, consistent with Croom et al.



The right panel of Figure~\ref{fig:b_z} displays the ratio of the quasar bias measured by the 2dF and the SDSS-II surveys to the value of our fitted function. 
Our results appear again to be compatible with former analyses of quasar clustering. We also stress that, with only one fifth of its final statistic, the eBOSS quasar sample already provides the most accurate measurement of the quasar bias in the redshift range $0.9 < z < 2.2$.

\begin{table}
\centering
\begin{tabular}{|cccccc|} \hline
$z_{min}$ & $z_{max}$ & $z_{eff}$ & $N_{\rm QSO}$ & $b_Q$ & $\chi^2$ (7 \rm d.o.f.) \\  \hline
0.9 & 1.2 & 1.06 & 13,594 & $ 1.75 \pm 0.08$ & $9.9$  \\
1.2 & 1.5 & 1.35 & 17,696 & $ 2.06 \pm 0.08$ & $1.7$  \\
1.5 & 1.8 & 1.65 & 17,907 & $ 2.57 \pm 0.09$ & $2.1$  \\
1.8 & 2.2 & 1.99 & 19,575 & $ 3.03 \pm 0.11$ & $8.5$ \\ \hline
0.9 & 2.2 & 1.55 & 68,772 & $ 2.43 \pm 0.05$ & $4.5$ \\
\hline
\end{tabular}
\caption{Fit of bias over $10<r<85$ $h^{-1}$Mpc in various redshift bins.}
\label{tab:b_z}
\end{table}






\section{The halo mass and the duty cycle of eBOSS quasars} 
\label{sec:HM}
We now discuss possible implications of our measurements of the clustering of $\sim 70{,}000$
eBOSS quasars for the host dark matter haloes of quasars at $1 \lesssim z \lesssim 2$. We quantify 
the activity of eBOSS quasars by calculating their duty cycle, through which 
the halo mass of a population of quasars can be linked to their luminosity.  

\subsection{Characteristic Halo Mass}

Quasars are biased tracers of underlying dark matter 
\citep[e.g][]{mw01} and the fact that 
more massive haloes have higher clustering bias \citep{kai84, tin10}, has been 
used as the basis for constraining the mass of the dark matter haloes that host quasars
\citep[e.g.][]{col89,ha01,mw01,wy05,ef15}. Here, we follow a similar approach to \citet{ef15},
who constrained the dark matter halo mass and duty cycle of $\sim 75{,}000$ quasars at 
$z\sim 2.5$ in the final release of the BOSS survey. 

We adopt  parameters such as to get a $\Delta = 200$ matter overdensity in the 
formalism of \citet{tin10} in order to calculate the minimum halo mass, 
$\mhmin$, and the characteristic halo mass, $\mhbar$, of our quasars.
We apply this approach to quasars in our main 
sample, and in each of our four redshift subsamples (detailed in Table\,\ref{tabinit}). 
In this formalism, $\mhbar$ is the characteristic halo mass that corresponds to 
our measured clustering bias, i.e.~$b(\mhbar)=b_{\rm Q}$, and $\mhmin$ is 
the minimum halo mass that bounds the range of haloes that correspond to the observed 
clustering bias, i.e. $b(M>\mhmin)=b_Q$, with 

\begin{equation}\label{eqn:bbar}
b(M > M_{h,{\rm min}}) \equiv \frac{\int_{M_{h,{\rm min}}}^{\infty} 
\frac{{\rm d}n}{{\rm d}M} b(M) {\rm d}M} {\int_{M_{h,{\rm min}}}^{\infty} \frac{{\rm d}n}{{\rm d}M} {\rm d}M}~,
\end{equation}

\noindent where the halo masses above $\mhmin$ are weighted by the halo abundance 
${\rm d}n/{\rm d}M$ in the halo mass function as determined by \citet{tin08}.

The assumption of a lower limit $\mhmin$ in Eqn.\,\ref{eqn:bbar}, suggests that haloes with $M 
<\mhmin$ can only host quasars that are less luminous than the least luminous 
quasar in our sample. This interpretation can be tested for consistency by checking whether 
quasar clustering is luminosity-dependent. For example, \citet{ef15} found that
the assumption of a scatter-less monotonic relation 
between halo mass and quasar luminosity failed to describe the observed lack of 
luminosity dependence for the clustering of BOSS quasars at $z\sim 2.4$. 
How quasar clustering varies with luminosity appears to be a subtle effect. Categorically 
detecting whether different luminosity quasars are hosted by different mass haloes 
will therefore require very precise measurements of quasar clustering.
Constraining any 
luminosity dependence to quasar clustering is a topic where eBOSS could make gains, 
given its expected unprecedentedly large sample of homogeneously selected quasars.

Prior to eBOSS, the most extensive wide-area spectroscopic quasar surveys at 
$z\sim1.5$ that were used for clustering analyses, were the 2dFQSO 
redshift survey \citep[2QZ;]{cr04,cr05} and the SDSS-DR5 quasar survey 
\citep[DRQ5;]{sch07,ro09}. Restricting to uniformly selected quasars over the redshift range $0.9 < z < 2.2$, these surveys
provided catalogs of $\sim20{,}000$--25{,}000 quasars with which to conduct clustering analyses. 
Projecting from SEQUELS, eBOSS is expected to spectroscopically confirm 
$\sim 70$\,quasars per deg$^2$ down to a limiting magnitude of $g<22$ over $\sim 7{,}500$\,deg$^2$, 
for a total sample of more than 500{,}000 uniformly selected quasars in the redshift
range $0.9 < z < 2.2$ \citep{Mye15}. 
The average magnitude of eBOSS quasars is $\sim$2.5 times fainter than that
of previous SDSS clustering samples, while covering a similar redshift range. Essentially, therefore, eBOSS will extend 
quasar clustering measurements by about a factor of 10 in luminosity. This unprecedented expansion of the dynamical 
range and number density of quasar samples will allow eBOSS to provide the highest statistical power yet to disentangle the  luminosity and redshift dependences of  quasar clustering.

Figure \ref{mzplot} shows $\mhmin$ and $\mhbar$ for our full (NGC+SGC) sample of 68{,}772 
quasars at $z\sim1.5$, as well as for our four redshift subsamples at 
$z=1.06,~ 1.35,~ 1.65$, and $1.99$. In addition, the 4th and 5th columns of Table\,\ref{tabmass} list
the $\mhbar$ and $\mhmin$ we derive for our four redshift subsamples as well as for our main  
sample. The errors on 
$\mhbar$ and $\mhmin$ are calculated from the confidence intervals for
the quasar biases that we derive from our clustering measurements.
These confidence intervals are projected through 
Eqn.\,\ref{eqn:bbar} at the mean redshift of each sample, using the \citet{tin08} halo mass function and
the appropriate values of $\mhmin$, in
order to derive a corresponding confidence interval in halo mass.

To illustrate how $\mhmin$ and $\mhbar$ change
over the redshift range that is covered by both BOSS ($z>2.2$) and eBOSS 
($z<2.2$), Figure \ref{mzplot} displays the measurements made by \citet{ef15} for BOSS quasars using the same 
formalism that we use here for eBOSS quasars. Figure \ref{mzplot} also includes
the same quantities estimated using the quasar clustering measurements 
from \citet{She07} at $z\sim3.1$ and $z\sim4.0$ and  
\citet{fr13} at $z\sim 2.4$. Note that the agreement between \citet{fr13} and
\citet{ef15} is not particularly surprising, as both measurements are made using
BOSS quasars. However, the reason for the extreme differences in  
halo mass measured by \citet{She07}, as compared to lower-redshift studies, remains debatable. \citet{She07} studied the 
clustering of a sample of $\sim 4000$ highly luminous quasars with a density of $\sim1$\,deg$^{-2}$
and measured quasar biases approaching $b_{\rm Q}\sim16$ at $z > 4$.
It is possible that there is a sharp change in the host halo mass of quasars that lie beyond 
the luminosity and redshift range of BOSS --- models in which quasars are triggered by major mergers
of gas-rich galaxies \citep[e.g.][]{Hop07}
do allow for evolutionary scenarios in which the clustering of luminous quasars
simply tracks the growth of the most massive haloes at $z > 3$.
Indeed, the duty cycle of $f_{\rm duty}\sim 1$ measured by \citet{She07} for
quasars at $z > 3$ implies that {\em all} rare supermassive haloes ($>10^{13}\,M_{\odot}$) host active 
black holes.

Previous authors \citep[e.g.][]{cr05} found convincing evidence that the bias of $z < 2.5$
quasars, at magnitudes of about $g < 21$, increases with redshift from $z\sim0.5$ to
$z\sim2.5$. This implies that the mass of the
haloes hosting quasars remains fairly constant at $z < 2.5$, because a higher bias can offset the 
fact that the characteristic mass of the average halo must dwindle at higher redshift (as structure
has had less time to grow).
By extension, if the bias of quasars were to remain constant at higher and higher redshift, this would
imply that the characteristic mass of the haloes hosting quasars was decreasing with redshift. Essentially
this dwindling host halo mass
was what was found by \citet{ef15} for BOSS quasars at $z > 2.5$, as is shown in Figure \ref{mzplot}.
Contrary to the flat $b_{\rm Q}(z)$, or dwindling host halo mass, measured for BOSS quasars at $z > 2.5$, 
the biases we measure for eBOSS quasars increase with redshift, implying that the characteristic host
halo mass of eBOSS quasars is roughly constant (again as shown in Figure \ref{mzplot}). This is in excellent agreement 
with the results of \citet{cr05}, who found a non-evolving halo mass
of $M = (3.0 \pm 1.6) \times 10^{12}\,h^{-1} M_{\odot}$ over $0.5 < z < 2.5$ for a smaller sample of quasars
that were slightly more luminous than those in our sample.

\subsection{Duty Cycle}

The length of duration of the quasar phase (the so-called ``duty cycle'') has been defined
in multiple slightly different ways in the literature. Here, we take the definition of the
duty cycle as the ratio of the number 
density of haloes that host black holes that are ``on'' (and thus observed as luminous quasars) 
to the full number of haloes that could host quasars within the luminosity range of 
our sample. As in \citet{ef15}, we compare the cumulative luminosity 
function of quasars over a range of luminosities to the cumulative space density
of haloes over the corresponding range of host halo masses \citep[e.g.][]{ha01,mw01}

\begin{equation}\label{eqn:fduty}
f_{\rm duty} = \frac{\int_{L_{\rm min}}^{L_{\rm max}} \Phi(L) {\rm d}L}{\int_{M_{\rm h,min}}^{\infty} \frac{{\rm d}n}{{\rm d}M} {\rm d}M}~,
\end{equation}

\noindent where the value of $\mhmin$ is set by the measured quasar bias (as in Eqn.\,\ref{eqn:bbar}),  
${\rm d}n/{\rm d}M$ is, again, taken from \citet{tin08}, and
$\Phi(L)$ is the quasar luminosity function. 
Note that we integrate our halo masses over the entire mass range from $\mhmin$
to infinity. Effectively, this reflects the extremely weak relationship between 
quasar clustering and quasar luminosity, by allowing the
quasars in our samples to be hosted by a limitless range of halo masses 
above $\mhmin$.
We adopt a recent quasar luminosity function from \citet{Pal16} that was derived using quasars in our redshift and luminosity ranges of interest.    
We use this luminosity function to calculate the 
space density of quasars in our samples (see the 3rd column in Table~\ref{tabmass}). Quasars targeted as part of 
eBOSS do not all receive a fiber for follow-up spectroscopy. Further, eBOSS is not
complete to {\em all} quasars in the Universe. Hence,
the observed number density of quasars listed in Table\,\ref{tabinit} should be lower than the
expected total space density of $0.9 < z < 2.2$ quasars at the flux 
limit of eBOSS, even if the \citet{Pal16} luminosity function is perfectly accurate. 

\begin{figure*}
\begin{center}
\includegraphics[angle=0,scale=0.35]{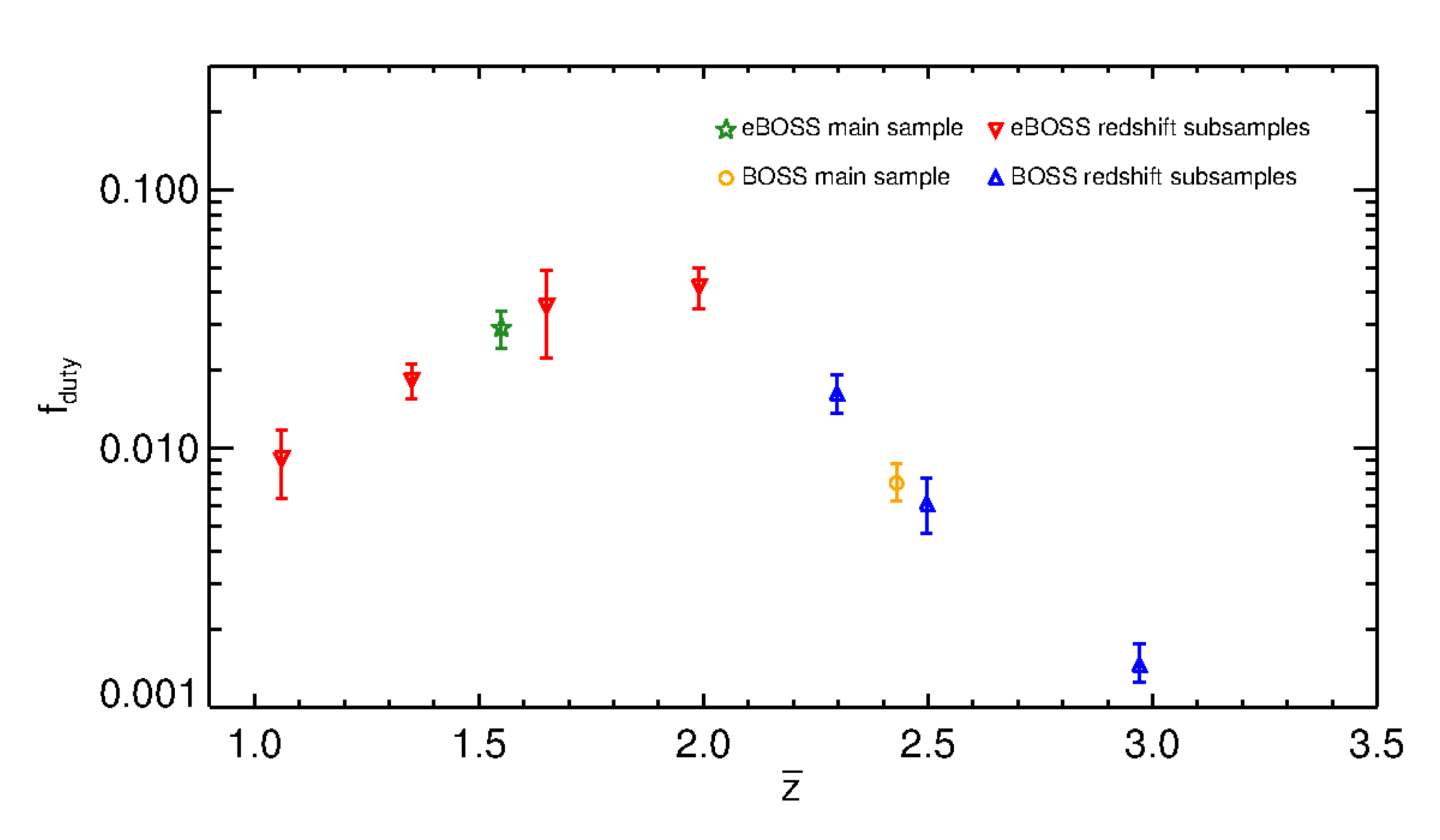}
\caption{The evolution of the duty cycle for our main sample (green star) and
 four redshift subsamples (red inverted triangles) calculated using Eqn.\,\ref{eqn:fduty}. See Table \ref{tabmass} for the 
$\fduty$ values and Table\,\ref{tabinit} for a summary of each sample's 
physical properties. Triangles depict values of $\fduty$ for BOSS quasars from
\citet{ef15} for their main sample (orange circle) and their three redshift subsamples (blue 
triangles).
}
\label{dcplot}
\end{center}
\end{figure*}

We display our calculated $\fduty$ values as a function of redshift in
Figure \ref{dcplot} and list the corresponding measurements in Table\,\ref{tabmass}.
We estimate errors on $\fduty$ by drawing sample values of the quasar bias from 
a Gaussian corresponding to the 68\% confidence interval around our measured $\pm 1\sigma$ errors on
$b_{\rm Q}$. We then calculate $\fduty$ for each sampled $b_{\rm Q}$ using
Eqn.\,\ref{eqn:bbar} and Eqn.\,\ref{eqn:fduty}, and hence derive the implied $\pm 1\sigma$ errors on $\fduty$.
Figure \ref{dcplot} compares our results to the similarly calculated 
$\fduty(z)$ of BOSS quasars at $z > 2.2$ from \citet{ef15}.
 
\begin{figure*}
\begin{center}
\includegraphics[angle=0,scale=0.35]{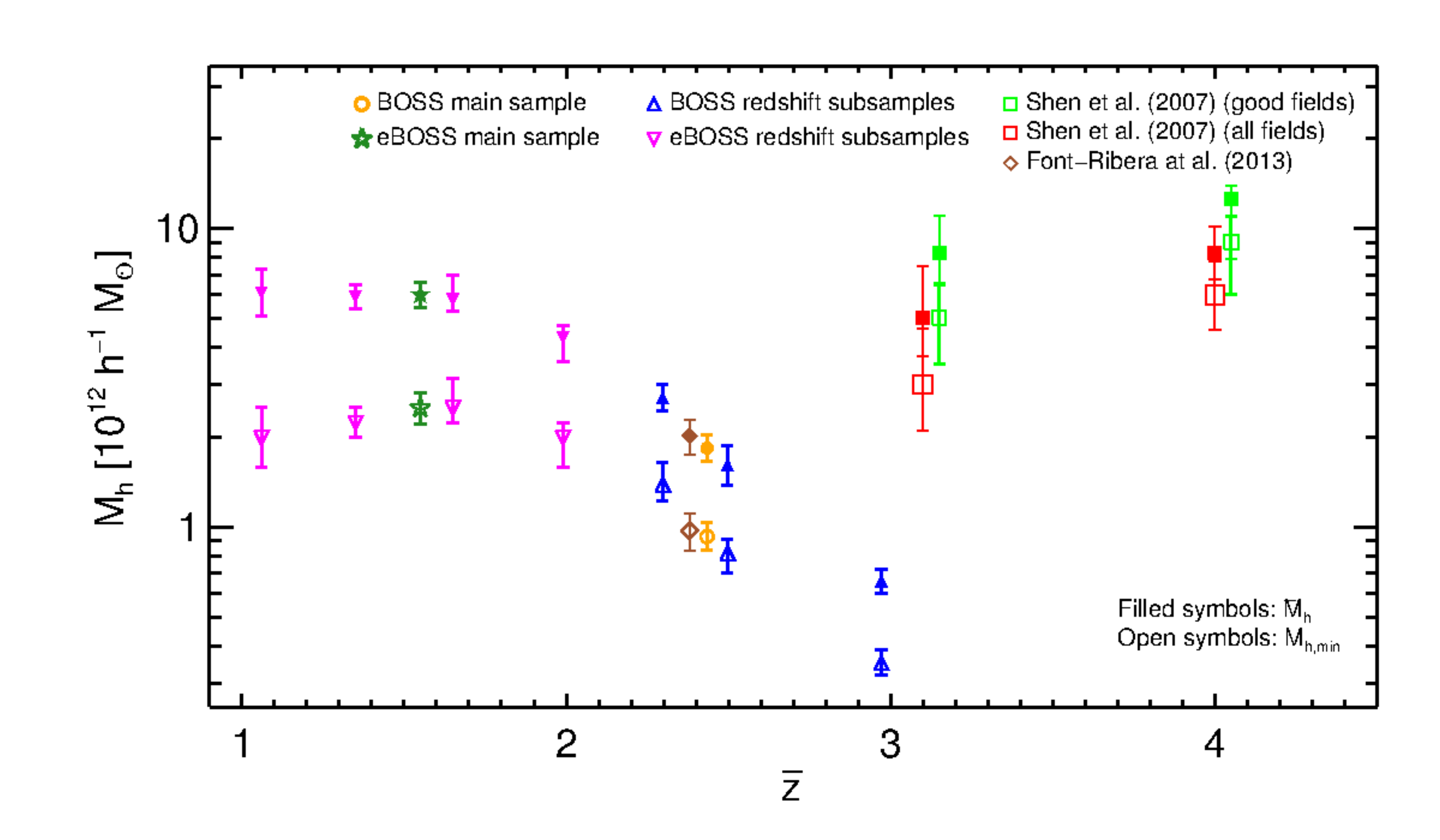}
\caption{The evolution of the minimum and characteristic halo mass (see also
Eqn.\,\ref{eqn:bbar} and Table \ref{tabmass}) for our full sample (green stars) and four 
redshift subsamples (pink inverted triangles).
Other points indicate the results for BOSS from \citet{ef15} for their main sample
(orange circle) and three redshift subsamples (blue triangles), from 
\citet[][the "good fields" points have been offset slightly for visual clarity]{She07}, and 
from \citet{fr13}. Results from previous works are based on their reported values of quasar bias, 
recalibrated to our chosen cosmology.}
\label{mzplot}
\end{center}
\end{figure*}

Under the assumption that there is effectively no link between the 
luminosity and clustering of quasars (i.e.\ the assumption that we used to derive $\fduty$), 
we can ignore the different luminosity ranges probed by BOSS 
and eBOSS and directly compare the host halo masses and duty cycles of BOSS and eBOSS quasars.
The almost flat $\mhbar(z)$ up until $z\sim1.8$ depicted in Figure \ref{mzplot}, implies that quasars reside 
in haloes of similar mass at $z\lesssim2$. Above $z\sim2$, the
characteristic mass of the haloes that host quasars appears to plummet, by almost a dex by $z\sim3$. 
Further, as listed in Table\,\ref{tabmass}, 
the measured duty cycle for eBOSS quasars at $\bar 
z\sim1.5$ is more than four times longer than for BOSS quasars at $\bar z\sim 2.5$.
It has long been known that the quasar population peaks in space density around
redshift 2--3 \citep[e.g.][]{Ric06}. We can interpret this peak as a physical manifestation

of a combination of the quasar duty cycle and the characteristic masses of quasar-hosting haloes.
As more massive haloes are rarer, $z\sim 2$--3 is a sweet-spot where duty cycles 
are large compared to host halo rarity. Below $z\sim2$ quasar-hosting haloes are equally as 
rare as they are at $z\sim2$ (because the
characteristic halo mass is unchanging) but the increasingly small duty cycle at lower redshifts 
implies that fewer of these haloes
host active quasars. in contrast, at $z\sim2$--3, the characteristic mass of quasar-hosting haloes drops, which implies
that quasar-hosting haloes are more common. This, however, is offset somewhat by a rapid reduction in the duty
cycle, which implies that at higher redshifts in the range $z\sim2$--3 fewer and 
fewer quasars are ``on'' in these increasingly more numerous haloes.

On the other hand, our assumption that there is absolutely no correlation between quasar
luminosity and host halo mass may break down under further scrutiny.
More sophisticated models that add scatter to the halo mass-luminosity relation 
\citep[e.g.][]{sha09} would then be needed to fully understand the interplay
between quasars and large-scale structure.  
The characteristic mass of the haloes that host quasars is an average
across the halo mass function (${\rm d}n/{\rm d}M$), so the fact that the characteristic mass stays
relatively constant between $z\sim 2$ and 
$z\sim 1$ could simply mean that the most massive haloes dominate this average.
A plausible scenario might be that less luminous quasars
inhabit a wide range of halo masses but more luminous quasars only reside in the
most massive haloes. At $z\sim1$, where we sample far down the quasar luminosity
function, we might then expect to see a wide range of halo masses, but the 
clustering signal would still be dominated by the most massive haloes. At $z\sim2$,
where our magnitude-limited sample
would shift to more luminous quasars, we would increasingly sample just higher-mass haloes.
In either case, at $z\sim1$ or at $z\sim2$ our clustering signal would only
reflect the clustering of high mass haloes. It is straightforward to interpret our measurements
under this alternative scenario. For example, 
Table\,\ref{tabmass} shows that quasars in our first redshift subsample at 
$0.9<z<1.2$ are the least luminous population, on average, among our four redshift subsamples, and that
there are also fewer of them. These
$0.9<z<1.2$ quasars have an $\mhmin$ that is somewhat smaller than the 2--3$\times$ more 
luminous population at $1.5 < z < 1.8$, but have an $\mhbar$ that is consistent. This could be interpreted to
be indicative of the less-luminous-than average $0.9<z<1.2$ quasars occupying the widest range of halo masses
in eBOSS but, also being less numerous, still having a clustering signal that is dominated by the most massive haloes.

Our sample of quasars is of insufficient size to detect any luminosity dependence to quasar clustering.
But, as was mentioned earlier in this section, a detailed study of the luminosity 
dependence of quasar clustering using the final eBOSS
sample of $\sim500{,}000$ quasars
remains an important and highly 
anticipated objective of the eBOSS survey. In addition, the quasars sampled by eBOSS 
overlap the Luminous Red
Galaxy and Emission Line Galaxy populations sampled by eBOSS around $0.7 \lesssim z \lesssim 1.0$. This
will provide a chance to cross correlate quasars with more-numerous galaxies \citep[as in, e.g.,][]{She13,Kro15} to try to
study the luminosity dependence of quasar clustering in narrow redshift bins near $z\sim0.8$.

\begin{table*}
\begin{center}
\begin{tabular}{cccc}
\hline
\hline
$\Delta z$ & $\Delta M_i$ & $N_{\rm qso}$& $n $ \\
    &   & & $(10^{-6}\,h^{-1}\rm Mpc)^{-3}$  \\
\hline
$0.9 \le z <1.2$ & $-22.05\le M_i\le-26.77$ &13594 &$13.94\pm0.27$ \\
$1.2 \le z <1.5$ & $-22.62\le M_i\le-27.33$ &17696 &$15.20\pm0.26$ \\
$1.5 \le z <1.8$ & $-22.97\le M_i\le-27.81$ &17907 &$13.98\pm0.27$\\
$1.8 \le z <2.2$ & $-23.49\le M_i\le-28.22$ &19575 &$10.87\pm0.30$\\
\hline
$0.9 \le z \le2.2$& $-22.82\le M_i\le-27.67$& 68772&$ 13.17\pm0.28$\\
\hline
\end{tabular}
\end{center}

\caption{The redshift limits, absolute $i$-magnitude range, total number of quasars (NGC+SGC) 
and space density in comoving coordinates for quasars in 
our main sample (final row) and redshift subsamples.}
\label{tabinit}
\end{table*}

\begin{table*}
\begin{center}
\begin{tabular}{cccccc}
\hline
\hline
$\Delta z$ & $\Delta L$ & $\Phi(L_{\rm min}< L <L_{\rm max})$ & $\mhmin$ & $\overline M_{\rm h}$  & $f_{\rm duty}$  \\
    &  ($10^{46}\,{\rm erg\,s^{-1}}$) & $(10^{-6}\,h^{-1}\rm Mpc)^{-3}$& \small ($10^{12}\,h^{-1} M_{\odot}$)& \small ($10^{12}\,h^{-1} M_{\odot}$) &  \\
\hline \\
$0.9 \le z <  1.2$ &  $0.04 \le L \le 2.96 $  & $ 16.96^{+1.54}_{-1.78} $& $ 1.99^{+0.52}_{-0.41}$ & $6.10^{+1.20}_{-1.00}  $ & $0.0091\pm0.0027 $  \\
\\
$1.2 \le z < 1.5$ &  $ 0.06\le L \le 4.94 $  & $ 23.69^{+2.46}_{-2.19} $& $ 2.24^{+0.27}_{-0.24}$ & $5.91^{+0.56}_{-0.51}  $ & $0.0183\pm0.0028 $  \\
\\
$1.5 \le z < 1.8 $&  $ 0.09\le L \le 7.68 $  & $ 29.37^{+2.91}_{-2.99} $& $ 2.51^{+0.65}_{-0.27}$ & $5.80^{+1.20}_{-0.51}  $ & $0.0355\pm0.0133 $  \\
\\
$1.8 \le z \le 2.2 $&  $ 0.14\le L \le 11.23  $  & $ 32.89^{+3.10}_{-3.56} $& $ 1.99^{+0.24}_{-0.41}$ & $4.33^{+0.42}_{-0.74} $ & $0.0422\pm0.0077 $  \\ 
\\
\hline 
\\
$0.9 \le z \le2.2 $&  $ 0.04\le L \le 11.23 $  & $ 26.82^{+2.11}_{-2.42} $& $ 2.51^{+0.31}_{-0.27}$ & $6.01^{+0.58}_{-0.58}  $ & $0.0292\pm0.0048 $  \\
\\
\hline
\end{tabular}
\end{center}

\caption{The first two columns display the redshift limits and luminosity range for our
main sample (final row) and redshift subsamples. The 3rd column lists the space density of quasars in the
given redshift and luminosity ranges, 
calculated using the combination of the Pure Luminosity Evolution (PLE) and the Luminosity and Density Evolution (LEDE) models for the luminosity function (PLE+LEDE) from \citet{Pal16}. The 4th and 5th columns display the minimum and 
the characteristic halo mass calculated at the average 
redshift of the sample (see Eqn.\,\ref{eqn:bbar}). The 6th column lists the
duty cycle, which is derived from $\mhmin$ and 
$\Phi$ (see Eqn.\,\ref{eqn:fduty}). $f_{\rm duty}$ is expressed as a fraction of 
the Hubble time (9.785 $h^{-1}\,{\rm Gyr}$ in our adopted cosmology). }
\label{tabmass}
\end{table*}

\section{Conclusion}
The first year of observation of the eBOSS survey provides 68,772 homogeneously selected quasars in the redshift range $0.9 < z < 2.2$, which represents the largest quasar sample ever obtained in this redshift range. We use this quasar sample to measure the quasar correlation function $\xi(r)$. We investigate various sources of systematic effects that might impact the measurement of $\xi(r)$, and find that the main contribution arises from inhomogeneities in the quasar target selection. We provide a weighting scheme that mitigates the important systematic effects, and we show that the resulting correlation function is much closer to zero on large scales

The measured correlation function is in  agreement with a linear $\Lambda$-CDM model in the range $10<r<85$ $h^{-1}$Mpc. We measure the quasar bias of our sample to be $b_{Q} = 2.45 \pm 0.05$, at $\bar{z} = 1.55$. Splitting our sample into four redshift slices provides the evolution of $b_{Q}$ with  redshift, and  confirms that $b_{Q}$ increases with $z$ in the studied redshift range. These results are compatible with  previous findings by the 2dF and SDSS-II surveys. It is also remarkable that, with only one fifth of the final sample, the eBOSS survey already provides the most accurate measurement of $b_{Q}(z)$ in the range $0.9 < z < 2.2$.


Adopting \citet{tin10}'s formalism for the the dark matter distribution and halo mass function, we calculate the minimum halo mass, $\mhmin$, and the characteristic halo mass, $\mhbar$, of our quasar sample.
We use a recent luminosity function that was derived using quasars in our redshift and luminosity ranges of interest \citep{Pal16} to measure the duty cycle of eBOSS quasars at $z\sim1.5$ and for subsamples of these quasars in four slices of redshift over $0.9<z<2.2$ to investigate the redshift evolution of $\mhmin$, $\mhbar$, and $\fduty$. We conduct our $\mhmin$, $\mhbar$, and $\fduty$ calculations under the assumption that there is weak to no connection between quasar clustering and quasar luminosity. This assumption allowed us to compare the same calculations for BOSS quasars at $z>2.2$ to our measurements for much fainter quasars at $z<2.2$. 

We find that the characteristic mass of haloes hosting quasars remains relatively constant at $z<2.2$. This finding is in agreement with the non-evolving halo mass of quasars over $0.5<z<2.2$ found by \citet{cr05}. Our result is also in accord with the dwindling halo mass found for BOSS quasars at $z>2.2$ by \citet{ef15} as the structures have more time to grow at higher redshifts than at $z<2$ (see Fig.\,\ref{mzplot}). We find the duty cycle of eBOSS quasars at $\bar z\sim 1.5$ to be more than four times longer than  that of BOSS quasars at $\bar z\sim 2.5$. Combining the duty cycles of BOSS and eBOSS quasars in Fig.\,\ref{dcplot}, we interpret the observed peak at the quasar duty cycle around $z\sim2$ as a physical manifestation of having fewer quasars that are ``on'' at $z\sim 2-3$.
The average luminosity of eBOSS quasars at $0.9<z<1.2$ in our sample is 2-3 times less than quasars at $1.5<z<1.8$ 
and they appear to occupy a wider range of halo masses with smaller $\mhmin$ compared to quasars at $1.5<z<1.8$ (see Table\,\ref{tabmass}). Nevertheless, the clustering signal for both sets of quasars is dominated by the rare most massive halos in their occupied range of halo masses. The size of our current sample of quasars in the first year of eBOSS is insufficient to detect any luminosity dependence to quasar clustering. Whether quasar clustering is luminosity-dependent will be further  investigated with the final sample of $\sim$500{,}000 eBOSS quasars.

\section*{Acknowledgements}

Funding for the Sloan Digital Sky Survey IV has been provided by the
Alfred P. Sloan Foundation, the U.S. Department of Energy Office of
Science, and the Participating Institutions. SDSS acknowledges
support and resources from the Center for High-Performance Computing at
the University of Utah. The SDSS web site is www.sdss.org.

SDSS is managed by the Astrophysical Research Consortium for the Participating Institutions of the SDSS Collaboration including the Brazilian Participation Group, the Carnegie Institution for Science, Carnegie Mellon University, the Chilean Participation Group, the French Participation Group, Harvard-Smithsonian Center for Astrophysics, Instituto de Astrof\'isica de Canarias, The Johns Hopkins University, Kavli Institute for the Physics and Mathematics of the Universe (IPMU) / University of Tokyo, Lawrence Berkeley National Laboratory, Leibniz Institut f\"ur Astrophysik Potsdam (AIP), Max-Planck-Institut f\"ur Astronomie (MPIA Heidelberg), Max-Planck-Institut f\"ur Astrophysik (MPA Garching), Max-Planck-Institut f\"ur Extraterrestrische Physik (MPE), National Astronomical Observatories of China, New Mexico State University, New York University, University of Notre Dame, Observat\'orio Nacional / MCTI, The Ohio State University, Pennsylvania State University, Shanghai Astronomical Observatory, United Kingdom Participation Group, Universidad Nacional Aut\'onoma de M\'exico, University of Arizona, University of Colorado Boulder, University of Oxford, University of Portsmouth, University of Utah, University of Virginia, University of Washington, University of Wisconsin, Vanderbilt University, and Yale University.

\bibliographystyle{unsrtnat_arxiv}
\bibliography{biblio}	

\end{document}